\documentclass[journal]{IEEEtran}
\ifCLASSINFOpdf
\else

\fi
\usepackage[utf8]{inputenc}
\usepackage{graphicx}
\usepackage{tabularx,booktabs}
\usepackage[export]{adjustbox}
\usepackage[margin=0.5in]{geometry}
\usepackage{amsmath,cite}
\usepackage{amssymb}
\usepackage{gensymb}
\usepackage{bbm}
\usepackage{subfig}
\usepackage{listings}
\usepackage{mathtools}
\newtheorem{lemma}{\textbf{Lemma}}
\newtheorem{corollary}{\textbf{Corollary}}
\newtheorem{theorem}{\textbf{Theorem}}
\newtheorem{approximation}{\textbf{Approximation}}
\newtheorem{remark}{\textbf{Remark}}
\usepackage{pgfplots}
  \pgfplotsset{compat=newest}

\usepackage{wasysym} 
\usepackage{graphicx}
\newcommand\tinyvarhexagon{\vcenter{\hbox{\scalebox{0.7}{$\varhexagon$}}}}

\usepackage{tikz}
\usetikzlibrary{patterns}

\DeclarePairedDelimiter\floor{\lfloor}{\rfloor}
\usepackage{tikz}
\newcommand{\slfrac}[2]{\left.#1\middle/#2\right.}

\usepackage{color,soul}

\begin{document}
\title{Data Aggregation in Synchronous Large-scale IoT Networks: Granularity, Reliability, and Delay Tradeoffs}
\author{Yasser~Nabil, Hesham ElSawy, Suhail Al-Dharrab, Hassan Mostafa, and Hussein Attia
\thanks{Y. Nabil and H. Mostafa are with the Department of Electronics and Communication Engineering, Cairo University, Giza 12613, Egypt,  e-mails: \texttt{\{yassernabil3, hassanmostafahassan\}@gmail.com}. H. Mostafa is also with the University of Science and technology, Nanotechnology and Nanoelectronics Program, Zewail City of Science and Technology, Giza 12578, Egypt. \\
H.\ ElSawy, S. Al-Dharrab, and H. Attia are with the Electrical Engineering Department, King Fahd University of Petroleum \& Minerals (KFUPM), Dhahran 31261, Saudi Arabia, e-mails: \texttt{\{hesham.elsawy, suhaild, hattia\}@kfupm.edu.sa}.}}
\maketitle

\begin{abstract}
This paper studies data aggregation in large-scale regularly deployed Internet of Things (IoT) networks, where devices generate synchronized time-triggered traffic (e.g., measurements or updates). The data granularity, in terms of information content and temporal resolution, is parameterized by the sizes of the generated packets and the duty cycle of packet generation. The generated data packets at the devices are aggregated through static terrestrial gateways. Universal frequency reuse is adopted across all gateways and randomized scheduling is utilized for the IoT devices associated with each gateway. Such network model finds applications in environmental sensing, precision agriculture, and geological seismic sensing to name a few. To this end, we develop a novel spatiotemporal mathematical model to characterize the interplay between data granularity, transmission reliability, and delay. The developed model accounts for several IoT design parameters, which include packet sizes, generation duty cycle, devices and gateways spatial densities, transmission rate adaptation, power control, and antenna directivity. For tractable analysis, we propose two accurate approximations, based on the Poisson point process, to characterize the signal-to-interference-plus-noise-ratio (SINR) based  transmission reliability. For the delay analysis, we propose a phase-type arrival/departure (PH/PH/1) queueing model that accounts for packet generation, transmission scheduling, and rate-sensitive SINR-based packet departure. The developed model is utilized to obtain the optimal transmission rate for the IoT devices that minimizes delay.  The numerical results delineate the joint feasibility range of packet sizes and inter-arrival times for data aggregation and reveal significant gains when deploying directional antennas.

\end{abstract}


\IEEEpeerreviewmaketitle

\section{Introduction}

\IEEEPARstart{T}{he} Internet of Things (IoT) provides ubiquitous wireless connectivity to billions of devices, which is an enabler for numerous applications that could enhance our world~\cite{Fuqaha2015, Palattella2016}. Big data aggregation is a core IoT application, which provides situational awareness for smart and automated environmental control. IoT networks are typically constituted of large groups of devices that are spatially distributed over wide geographical areas. Such IoT devices are equipped with multiple sensors to measure, track, and report some physical phenomena. Owing to their dense spatial deployment, such IoT devices will generate enormous amounts of data~\cite{first_mile}. Hence, the data granularity has to be carefully adjusted to balance the tradeoff between the generated wireless traffic intensity and the required accuracy of the  underlying application. In general, more information content (e.g., higher resolution per measurement and/or more sensors per device) necessitates more encoding bits. For a fixed source coding scheme, larger packet sizes imply more informative data. Furthermore, reducing the period between successive measurements and/or updates improves the time resolution of the data. However, the improved data granularity, in terms of the information content and time resolution, comes at the expenses of prolonged wireless transmission, increased energy consumption, and aggravated mutual interference within the network.  
To balance the critical tradeoff between data granularity, transmission reliability, and packet delay, it is required to account for the interplay between several network parameters and key performance indicators such as the traffic intensity, devices density, mutual interference, transmission success probability, and delay. Such interplay can be captured by spatiotemporal models, which are based on stochastic geometry and queueing theory~\cite{first_mile, tractable_delay}. In particular, the IoT devices are considered as a network of spatially interacting queues to characterize the effect of traffic generation on the induced mutual interference between active IoT devices. Packets are generated and stored at each IoT device to be transmitted to their intended receivers. Such transmissions create mutual interference in the network. On the other hand, packet departures require scheduled transmission grants that attain a rate-dependent signal-to-interference-plus-noise-ratio (SINR) threshold at the intended receiver. Using spatiotemporal models, the queues stability of the interacting devices in large-scale networks is studied in~\cite{gharbieh2017spatiotemporal, stability_alammouri, Stability_Haenggi, Chisci, SpatialBD, Deng_RA}. The delays imposed by different scheduling and random access schemes are quantified and compared in~\cite{paradox, Robin_IoT}. The self-sustainability of energy-harvesting IoT devices is characterized in \cite{Gharbieh_harvest, fatma_harvest}. Delay analysis for prioritized traffic in IoT networks is provided in~\cite{priority_Emara, priority_nardilli}. Rate adaptation to reduce delay is proposed in~\cite{el2020rate, Haenggi_rateless}. The impact of massive antenna systems on the performance of IoT networks is quantified in~\cite{Massive_iot}. Age of information for IoT traffic is characterized in \cite{emara2020spatiotemporal, Arafa_Iot, Harpreet_AoI}. The impact of transmission deadlines on the performance of IoT networks is studied in~\cite{elsawy2020characterizing}.

Based on the underlying application, IoT networks can be deployed in different spatial setups to handle different forms of traffic patterns. For large-scale IoT networks, most of the spatiotemporal models in the literature are devoted towards stochastic spatial topologies and randomized event-triggered traffic patterns~\cite{Stability_Haenggi, Chisci, SpatialBD, fatma_harvest, priority_Emara, el2020rate}. Few studies develop spatiotemporal models for asynchronous time-triggered traffic in IoT networks with stochastic spatial topologies~\cite{elsawy2020characterizing, emara2020spatiotemporal}. However, large-scale regular IoT network topologies with synchronous time-triggered traffic are overlooked. {In some applications such as ecological monitoring, smart agriculture, and mining/geophysical seismic surveys, the IoT devices are usually static and are regularly spaced from one another.  This would give a better estimation of the measured physical parameter (e.g., humidity, temperature, moisture, subsurface geological structure, etc.) across the field of interest. However, such large-scale regular IoT  networks are not well investigated in the literature, which can be attributed to its less tractability when compared to randomized network deployment as stated in \cite[Chapter~2]{haenggi2009interference}}. Furthermore, we are not aware of a mathematical model that characterizes the tradeoff between data granularity, transmission reliability, and delay in large-scale IoT networks.
To the best of the authors' knowledge, this paper is the first to model and assess large-scale IoT networks with grid (i.e., regular) spatial topology and synchronous time-triggered traffic, where the underlying traffic model accounts for the data granularity. Such regular network topology with data granularity-aware synchronous traffic finds applications in different domains of ecological monitoring, environmental sensing \cite{sevegnani2018modelling}, smart agriculture \cite{friha2021internet}, wildfire detection \cite{pandey2018modeling}, and mining/geophysical seismic surveys \cite{reddy2019wireless}, \cite{Attia_2020}.

This paper introduces a spatiotemporal model for large-scale grid-based IoT networks with synchronous periodic traffic. The data generated at the IoT devices are aggregated via static terrestrial gateways. A phase (PH) type arrival process at the queue of each device is utilized to account for the data generation in terms of quantity (i.e., bits per packet) and frequency (i.e., packets generation duty cycle). The packet departures are modeled using another PH-type process that captures the scheduling and rate-aware transmission success probability as a function of the attained SINR at the intended receiver. For tractability, the SINR distribution of the typical uplink connection is obtained by approximate Poisson point process (PPP) analysis. In particular, we propose two accurate PPP models to characterize the SINR distribution in the considered grid-based network. The developed model accounts for the impact of inter-device spacing, packet sizes and generation duty cycle, antenna directivity and radiation pattern, and transmission rate on the data aggregation reliability and packet delay. The developed model reveals an optimal transmission rate that minimizes delay. The numerical results demonstrate significant gains for directional antennas in terms of transmission reliability and delay. The results also delineate the joint range of packet sizes and inter-arrival times for feasible data aggregation.

\section{System Model}\label{s_g}

\subsection{Spatial and Temporal Models}
 
 From the spatial perspective, an infinite grid topology is considered to account for a large-scale IoT deployment. The IoT devices are distributed on parallel lines with equal inter-line separation of $\Delta y$. On each line, the IoT devices are placed with equal inter-device spacing of $\Delta x$. Without loss of generality, we assume that $\Delta y \geq \Delta x$, otherwise, we rotate the network and switch notations. Static terrestrial gateways are regularly placed at midpoints between the parallel lines to collect the data generated from the IoT devices.
 
 Assuming that each gateway has a communication range of $R = n \Delta y$ for $n \in \mathbb{N}$, the gateways are placed according to a hexagonal grid to provide comprehensive coverage (i.e., with no gaps) for the IoT devices.  Let $\mathcal{N}_G$ denote the number of IoT devices served by each gateway, then using simple trigonometric analysis we have 
 \begin{equation}\label{nn1}
    \mathcal{N}_G= 2 \sum_{i=0}^{Y-1}\floor*{ {\frac{2R- (2 i + 1) \Delta y/\sqrt{3}}{\Delta x}}+\frac{1}{2}},
 \end{equation}
  where $\floor*{\cdot}$ is the floor operator and $Y$ is the number of lines passing through one half of the hexagon. From the hexagon geometry, $Y$ is given by 
  \begin{equation}\label{nn2}
    Y =  \floor*{\frac{\sqrt{3}R}{2 \Delta y} + \frac{1}{2}}.
 \end{equation}
  
\begin{figure}[t]
\centering
  \includegraphics[width=0.4\textwidth]{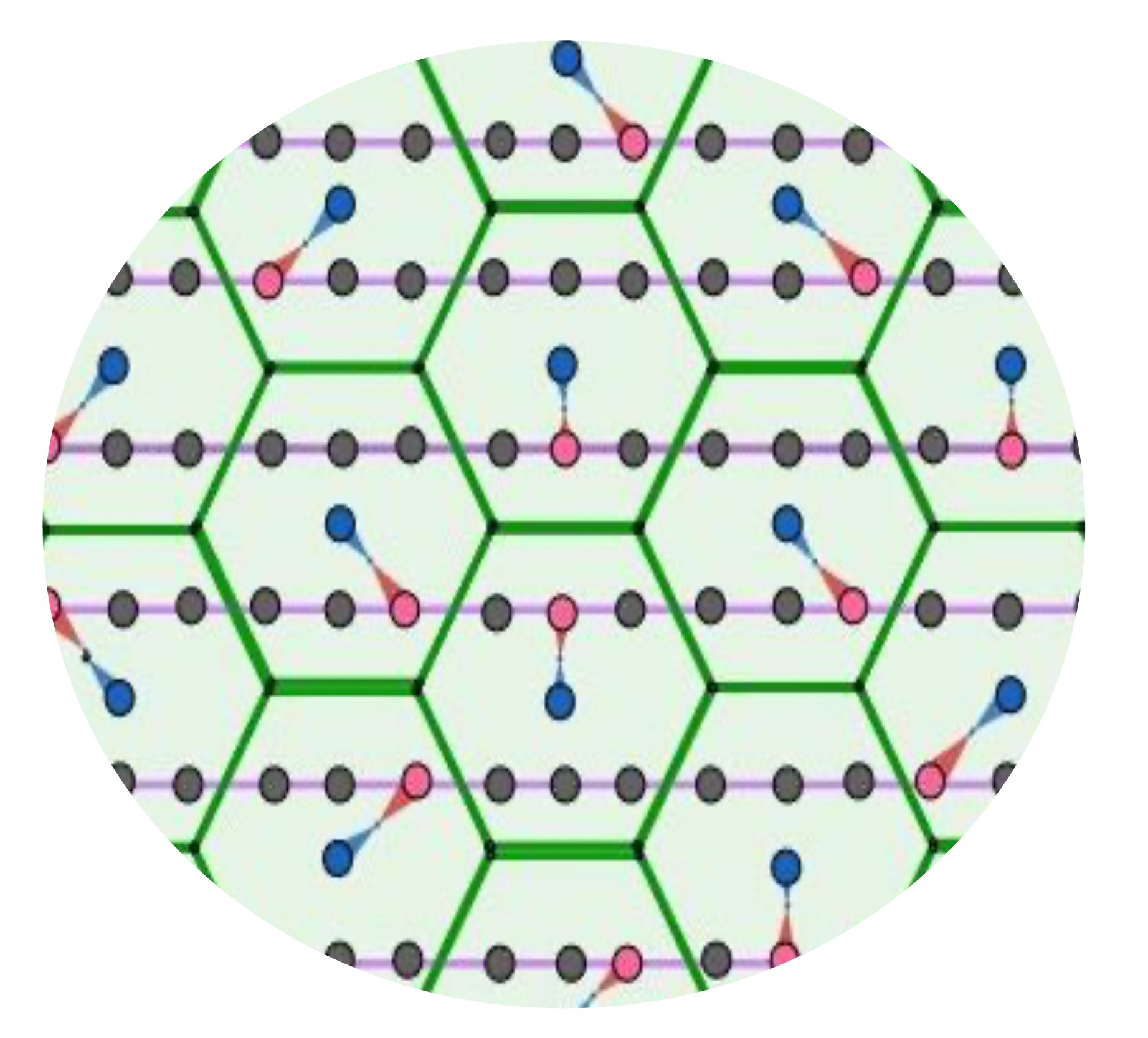}
    \caption{ {Network spatial model for $Y=1$ and $\mathcal{N}_G=6$. Only one device is active in each cell every time-slot, with perfectly aligned antennas for the intended link.}}
    \label{Spatial}
\end{figure}

  Using \eqref{nn1} and \eqref{nn2}, the gateway communication range $R$ can be adjusted to control the number of IoT devices $\mathcal{N}_G$ served by each gateway. An example of an IoT network realization is shown in Fig.~\ref{Spatial}.

From the temporal perspective, a time slotted system with a slot duration of $T_s$ is considered. The synchronous time-triggered traffic is represented via a deterministic arrival process of packets with size $L$ bits and duty cycle of $T_r \gg T_s$ seconds. All IoT devices have synchronized duty cycles for packet generation. The data granularity is parameterized by the parameters $L$ and $T_r$, which control, respectively, the information content and time resolution of the generated data. Larger values of $L$ imply more informative and/or higher precision data. Furthermore, low duty cycle $T_r$ entails more frequent packet generation, and hence, better data temporal resolution.   

\subsection{Network Model}

Uplink transmissions from the IoT devices to the gateways are subject to propagation losses and multipath fading. The signal power decays as a function of the propagation distance $r$ as $r^{-\eta}$, where $r$ is measured in meters and $\eta>2$ is the path-loss exponent. We assume a quasi-static multipath Rayleigh fading, where the channel gains remain constant during the time slot duration $T_s$ but randomly change across different time slots. Furthermore, the channel gains are independent of the devices' location and time slot indices. All power fading gains in the network are modeled as independent and identically distributed (i.i.d.) unit mean exponential random variables.

\begin{figure}[!h]
\centering
  \includegraphics[width=0.35\textwidth]{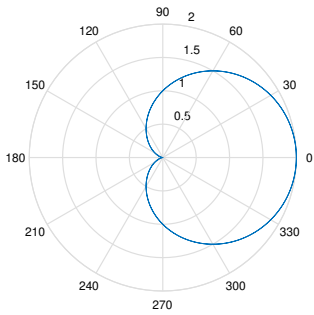}
    \caption{Antenna radiation pattern for $b=1$ and $n=1$.}
    \label{radiation}
\end{figure}

Power control and directive antennas are utilized to improve transmission reliability and reduce delay. The IoT devices adopt path-loss inversion power control to compensate for propagation losses. A device at distance $r$ from its serving gateway transmits with power $P_r=\rho r^{\eta}$. Hence, all devices maintain an average signal power of $\rho$ at their intended gateways irrespective of their locations. Equipping gateways and IoT devices with directive antennas improves the received signal quality and reduces mutual interference. Following \cite{georgiou2015directional, balanis2016antenna}, we utilize the following gain function for the directive antennas
\begin{equation}\label{second}
G(\theta )=1+b \ \mathrm{cos}(n\theta )
\end{equation}
where $0\leq b\leq 1$ controls the beamwidth of the main lobes and $n \in \mathbb{N}$ determines the number of main lobes. The gain in \eqref{second} represents the radiation intensity in a given direction $\theta$ when compared to the same power radiated from an isotropic antenna. The antenna radiation pattern is depicted in Fig. \ref{radiation} for the case of a single main lobe.


Due to the static network topology, the transmission power and antenna orientation are fixed once determined, which simplifies the IoT network operation. However, directive antennas are generally larger and more expensive than omni-antennas. Furthermore, power control may impose faster battery depletion for devices located farther away from their serving gateways. Owing to the stringent size, complexity, and energy constraints of IoT devices, we quantify the added value of directive antennas and power control by benchmarking against omni-antennas and constant transmission power. From the antenna directivity perspective, we study the following cases 
\begin{itemize}
    \item \textit{Directional Gateways (D-GW) \& Directional Nodes (D-N)}: this scenario prioritizes performance over complexity. 
    \item \textit{Directional Gateways (D-GW) \&  Omni-directional Nodes (O-N)}:  this is a balanced scenario that adds complexity to the gateways and utilizes low-cost devices. 
    \item \textit{Omni-directional Gateways (O-GW) \&  Omni-directional Nodes (O-N)}: this is the resource-stringent scenario that sacrifices  performance for low-cost network implementation. 
\end{itemize}
Each of the aforementioned scenarios is studied with and without power control.

\subsection{Transmission Model}

\begin{figure}[t]
\centering
  \includegraphics[width=0.5\textwidth]{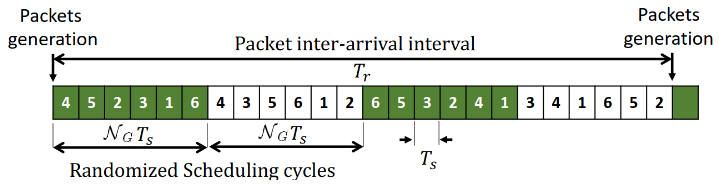}
    \caption{Randomized scheduling cycles for $\mathcal{N}_G=6$ and $T_r= 24\times T_s$.  The number within the time slot indicates the index of the scheduled device. The alternating colors highlight different scheduling cycles, where each device is granted $T_a=4$ uplink transmissions within the packets inter-arrival interval $T_r$.}
    \label{scheduling_cycle}
\end{figure}

Due to spectrum scarcity, a universal frequency reuse scheme of $W$ Hz is adopted across all gateways. Hence, there exists inter-cell interference between devices served by different gateways. However, an independent randomized scheduling scheme is utilized within each gateway to avoid overwhelming intra-cell interference.\footnote{From the delay perspective, it is shown in \cite{Robin_IoT} that randomized scheduling outperforms conventional round-robin schemes, which motivates our model.} To schedule all of its associated devices, each gateway utilizes a scheduling cycle of length $\mathcal{N}_G$ time slots. At the beginning of each cycle, each gateway generates and broadcasts a randomized schedule to its associated devices. Such transmission schedule randomly and independently changes across scheduling cycles and across different gateways. An example of four scheduling cycles for a gateway serving six devices is shown in Fig.~\ref{scheduling_cycle}.

From the device's point of view, each device in the network is granted intra-cell interference-free uplink access at a random time slot during each scheduling cycle. Within the traffic generation duty cycle of $T_r$, each device is granted $T_a = \frac{T_r}{\mathcal{N}_G T_s}$ uplink attempts to deliver the generated $L$ bits packet. However, due to fading and inter-cell interference, uplink transmission attempts are subject to decoding errors. To balance the tradeoff between transmission reliability and delay, each device divides its $L$ bits generated packet into $m < T_a$ smaller equally sized segments of size $\floor*{\frac{L}{m}}$ bits.\footnote{The number of segments $m$ should be less than $T_a$ for finite packet delay.} Smaller segments can be transmitted at lower rates, and hence higher reliability,  during the allocated time slots. Let $\mathcal{R}_m$ denote the transmission rate when the packet is segmented to $m$ equal segments, then we have 
\begin{equation}\label{rate}
\mathcal{R}_m=\frac{L}{m T_s}=\zeta W \log _{2}(1+\Xi_m)
\end{equation}
where $0 < \zeta \leq 1$ captures the gap between practical transmission rates when compared to Shannon’s capacity, and $\Xi_m$ is the SINR threshold required to
correctly decode the segment at the gateway when operating with the rate $\mathcal{R}_m$.  Solving \eqref{rate} for $\Xi_m$, a packet transmitted at rate $\mathcal{R}_m$ is successfully received at the gateway if and only if the intended SINR satisfies the following inequality 
\begin{equation}\label{rate2}
\text{SINR} \geq\Xi_m=2^{\frac{L}{m\zeta W T_s}}-1.
\end{equation}
Eq. \eqref{rate2} shows that higher $m$ (i.e., more segments with smaller sizes), requires lower decoding thresholds, and hence, has higher immunity to interference and noise. However, the higher reliability comes at the cost of more required transmissions per packet (i.e., to deliver all of its $m$ segments). 

\subsection{Methodology of Analysis}
To account for the interplay between data granularity, device density, and mutual interference, we model the IoT devices as a network of spatially interacting PH/PH/1 queues. In details, each device is represented by a queueing system with PH type arrival that captures the aforementioned packets generation with sizes $L$ and duty cycle $T_r$. Generated packets are stored within each IoT device to be transmitted to the intended gateway via a first-in-first-out (FIFO) discipline. A packet departure from each IoT device is modeled via another PH type process that accounts for the aforementioned i) randomized scheduling; ii) the utilized rate $\mathcal{R}_m$; and iii) SINR attained at the gateway. {Segments that  are  correctly  received  at  the  gateway  are  dropped  from  the device’s buffer. Otherwise, the transmission of the same segment is repeated until successful delivery}. Based on the aforementioned system model, the segment transmission success probability when operating at rate $\mathcal{R}_m$ can be expressed as 

\begin{align}\label{third}
&p_m = \mathbb{P}\{\text{SINR} >\Xi_m\}, 
\end{align}
such that 
\begin{align}\label{third33}
\!\!\text{SINR}\!=\!\frac{P_o h_o G(\theta_{T_o}) G(\theta_{R_o}) r_o^{-\eta}}{\sigma^2\!+\!\sum\limits_{i=1}^{\infty}\sum\limits_{j\in\Psi_i}\!\!\! \mathbbm{1}_{\{v_{ij} \text{ is active}\}}  P_{ij} h_{ij} G(\theta_{T_{ij}}) G(\theta_{R_{ij}}) \|v_{ij}\|^{-\eta}},
\end{align}
where $P_o$, $h_o$, $\theta_{T_o}$, $\theta_{R_o}$ and $r_o$ are, respectively, the transmission (Tx) power, fading gain, Tx antenna orientation, receiver (Rx) antenna orientation, and distance of the intended link. Assuming perfectly aligned antennas for the intended link implies that $\theta_{T_o} = \theta_{R_o} = 0$, and hence, $G(\theta_{T_o})\times G(\theta_{R_o}) = 4$. The noise power is denoted as $\sigma^2$. The set $\Psi_i$ encompasses the locations $v_{ij} \in \mathbb{R}^2$ of all potential interfering devices on each line of the gird network. The parameters of each interfering device in $\Psi_i$ are $P_{ij}$, $h_{ij}$, $\theta_{T_{ij}}$, and $\theta_{R_{ij}}$ which denote, respectively, the Tx power, interfering channel gain, interfering device antenna orientation with respect to (w.r.t.) the test gateway, and the test gateway Rx antenna orientation w.r.t. the interfering link. The interfering link distance is given by $\|v_{ij}\|$, where $\|\cdot\|$ denotes the Euclidean norm. The probabilistic activation of each interfering link is captured via the indicator function $\mathbbm{1}_{\{\cdot\}}$, which takes the value one if the statement $\{\cdot\}$ is true and zero otherwise.  

For organized exposition, we divide the spatiotemporal analysis into two main sections. Section~\ref{sp_m} presents the spatial analysis to characterize the SINR and obtain the segment transmission success probability in \eqref{third}. For tractability, we propose two PPP-based approximations for \eqref{third}, which are validated in Section~\ref{rs}. The transmission rate and delay analysis are then presented in Section~\ref{qm}, which are based on queueing theory. It is worth emphasizing that the packet departures in queueing theory analysis are SINR-aware, and hence, are based on the  segment transmission success probability obtained via the spatial analysis in Section~\ref{sp_m}.


\section{Spatial Model and Analysis}\label{sp_m}

This section characterizes the rate-aware uplink transmission success probability of a segment, hereafter denoted as success probability, given in \eqref{third}.  Without loss in generality, we evaluate the success probability at a test gateway located at an arbitrary origin. Due to the regular and symmetric network topology, the test gateway is typical and represents the success probability at all gateways in the network. Having said that, the SINR numerator in \eqref{third33} represents the intended signal power from one of the devices served by the test gateway. The interference term in the denominator of \eqref{third33} has an indicator function that accounts for the randomized scheduling for the devices served by other gateways. To find the density of interfering devices, we first recall that the inter-line separation is $\Delta y$ and the per-line inter-device separation is $\Delta x$. Hence, the total devices density is $\lambda_t=\frac{1}{\Delta x \Delta y}$ device per unit area. By virtue of randomized scheduling, only one of the $ \mathcal{N}_G$ devices becomes active per gateway at a given time slot. Hence, the density of interfering devices is
\begin{equation} \label{int_loc}
\lambda_a = \frac{1}{ \mathcal{N}_G}\times \lambda_t = \frac{1}{ \mathcal{N}_G \Delta x \Delta y}.   
\end{equation}

The interfering devices with the density in \eqref{int_loc} follow a 2D discrete and correlated random locations (i.e., with potential positions on the original gird and one device per hexagon). Furthermore, the orientation of each interfering device is a function of its location to account for the antenna alignment between the device and its serving gateway. Accounting for the exact locations and orientations of the interfering devices in \eqref{third33} impedes the mathematical tractability of the analysis. Hence, to maintain tractability, we utilize the following approximations. In particular, for the devices' orientation, we propose the following approximation. 

\begin{approximation}[Uniform random orientation]
Interfering devices that are located within the distance range $\frac{\sqrt{3}R}{2} \leq r \leq {{3}R}$ from the test gateway are assumed to have uniform random orientation over the range of $\theta \in [\frac{\pi}{2},\frac{3 \pi}{2}]$. Interfering devices that are located farther than $r> {{3}R}$ from the test gateway are assumed to have uniform random orientation over the range of $\theta \in [0,2 \pi]$.
\label{app_orint}
\end{approximation}

\begin{remark}
Since the devices are oriented towards their serving gateways, the interfering devices located within the distance range $\frac{\sqrt{3}R}{2} \leq r \leq \sqrt{3}R$ will have opposite antenna orientation w.r.t. the test gateway. Hence, their orientation is assumed to be within the range of $\theta \in [\frac{\pi}{2},\frac{3 \pi}{2}]$. This is not the case for devices that are farther than $r> \sqrt{3}R$, and hence, the full orientation range $\theta \in [0,2 \pi]$ is considered. It is worth noting that Approximation~\ref{app_orint} is effective for the case of directional devices only.  In the case of omni-directional devices, the power is uniformly radiated in all directions, and hence, Approximation 1 does not apply. 
\end{remark}

For the locations of the interfering devices, we propose the following two alternative PPP-inspired approximations. 
 \begin{approximation}[Parallel 1-D PPPs]
 The interfering~devices are approximated with parallel homogeneous one-dimensional (1-D) PPPs with inter-line separation of $\Delta y$ and per-line intensity of $\lambda_{1D}= \frac{1}{\Delta x \mathcal{N}_G}$ device per unit length. This approximation relaxes the discretized locations of the interfering devices but still restricts their positions to the parallel lines. Owing to the fact that the density of lines is $\frac{1}{\Delta y}$, the total intensity of the interfering devices in this approximation is $ \frac{1}{\Delta x \Delta y \mathcal{N}_G}$, which matches the intensity of the exact case in \eqref{int_loc}. 
 \label{app_1DPPP}
\end{approximation}

\begin{figure}[t!]
\centering
   \subfloat[\label{intr_field} ]{%
      \includegraphics[ width=0.4\textwidth]{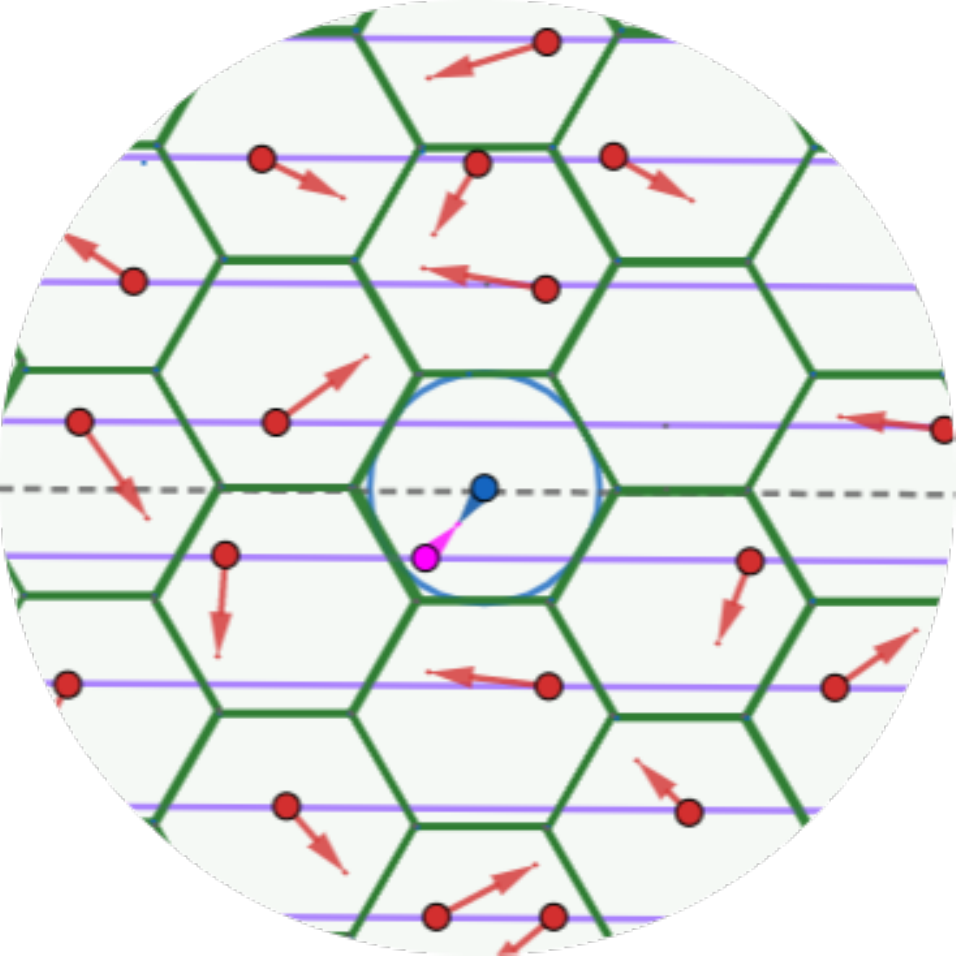}}
\hspace{\fill}
   \subfloat[\label{intr_field_arae}]{%
      \includegraphics[ width=0.4\textwidth]{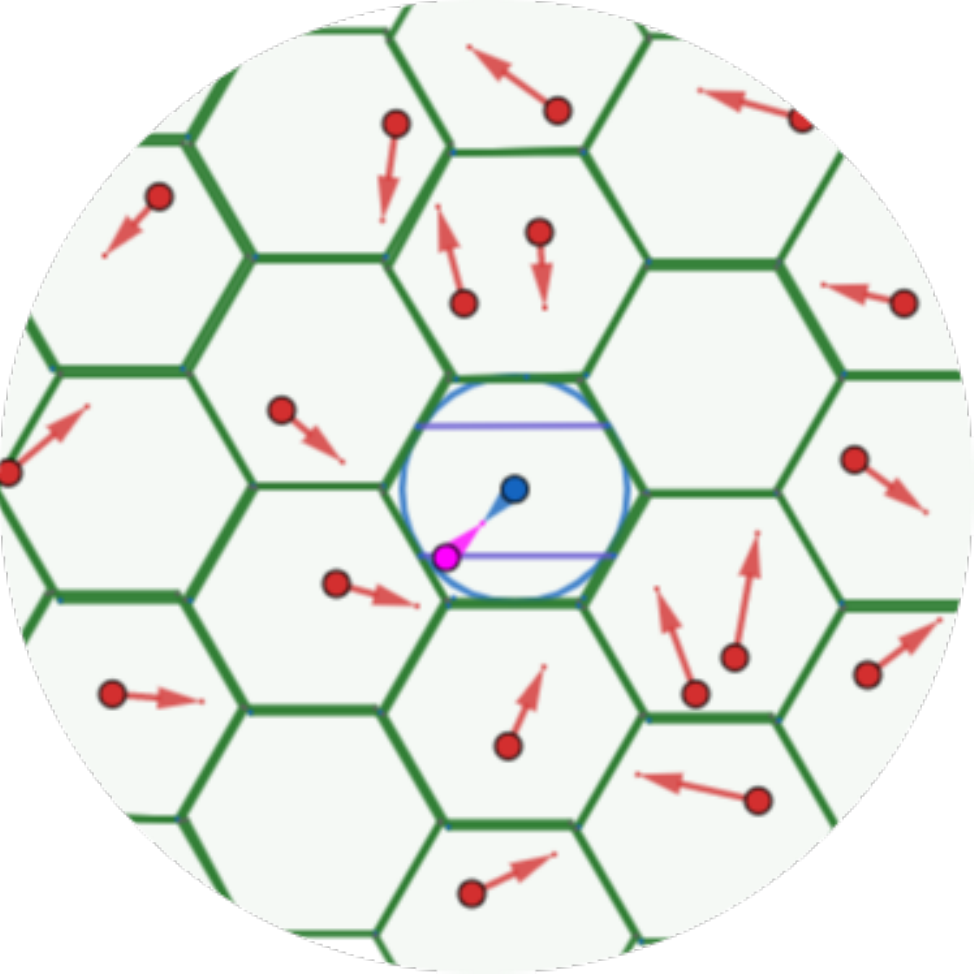}}\\
\caption{\label{workflow}  {Approximated system models  (a) 1D-PPPs    (b) 2D-PPP. Interfering devices are randomly distributed (a: over lines, b: over the whole area)  outside the interference free region with arrows representing the uniformly random antenna orientation for the interfering IoT devices.}}
\end{figure}

  \begin{approximation}[2-D  PPP]
  To further facilitate the analysis, the interfering devices are approximated  with a homogeneous two-dimensional (2-D) PPP $\Psi_{2D}$ with intensity $\lambda_{2D}= \frac{1}{\Delta x \Delta y \mathcal{N}_G}$ device per unit area. Such approximation relaxes all constraints on the relative locations of the interfering devices. However, it still matches the intensity of the exact case in \eqref{int_loc}. 
   \label{app_2DPPP}
\end{approximation}
 
 \begin{remark}
It is important to note that we always account for the intra-cell interference-free uplink access within the test gateway.  Consequently, the interfering devices are restricted outside a circular radius of $\frac{\sqrt{3}R}{2}$. Furthermore, the intensities of the interfering devices in Approximations~\ref{app_1DPPP}~and~\ref{app_2DPPP} match that of the exact model in \eqref{int_loc}. 
\end{remark}

 \begin{remark}
The main purpose of Approximations 1, 2, and 3 is to (i) relax the discretized locations of the interfering devices, (ii) remove the coupling between the device position and orientation, and (iii) relax the relative spatial correlations that enforce only one interfering device per gateway. Relaxing the discretized interfering devices' locations transforms infinite probabilistic combinatorial summations to simple stochastic geometry-based integrals.  {Letting the interfering IoT devices to have independent and identically distributed uniformly random orientations alleviates tedious mathematical complications that would impede the model tractability. The accuracy of such approximation is validated in Section V.} 
\end{remark}

Approximations~\ref{app_1DPPP}~and~\ref{app_2DPPP} are mutually exclusive and represent an intuitive tradeoff between accuracy and mathematical simplicity. Indeed the 1D-PPPs approximation is closer to the spatial topology of exact system model, and hence, is expected to provide better accuracy when compared to the 2D-PPP approximation. However, the 2D-PPP approximation offers simpler analysis when compared to the 1D-PPPs approximation. To quantify such accuracy/simplicity tradeoff, we study the following two scenarios.  In the first scenario, we utilize Approximation 1 for the devices' orientation along with Approximation 2 for the devices' location, which we denote hereafter 1D-PPPs. The second scenario, we utilize Approximation 1 for the devices' orientation along with Approximation 3 for the devices' locations, which we denote hereafter 2D-PPP.  A pictorial illustration of both scenarios is shown in Fig.~\ref{workflow}. Both approximations are validated in Section~\ref{rs}.

\subsection{1D-PPPs Model with Constant Tx Power}

This section utilizes Approximations~\ref{app_orint} and \ref{app_1DPPP} to characterize the success probability in \eqref{third}. Furthermore, we focus this section on the constant transmission power scheme. The effect of power control is studied later in Section~\ref{power_control_sec}. Following the common practice in stochastic geometry, we first find the Laplace transform (LT) of the aggregate interference seen at the test gateway.  

\begin{lemma} \label{lemm1}
For the D-GW and D-N scenario with constant transmission power $(P)$, the LT of the 1D-PPPs aggregate interference observed at the test gateway located at the origin can be expressed as \eqref{ee12} presented at the top of next page, where $ \psi_i(\cdot)= \text{max}(\Delta y(i+\frac{1}{2}),\cdot)$ and  $C(\gamma,i)$ is given by 
\begin{figure*}
\small
\begin{equation}\label{ee12}
\begin{aligned}[b]
\!\!\mathcal{L}_{I_{agg}} (s)\!=\!& \exp\left\{\!-\! \sum\limits_{i=0}^{\infty} \int\limits_{0}^{2\pi}\! \frac{1}{\pi^2 \Delta x  \mathcal{N}_G} \left( \int\limits_{\frac{\pi}{2}}^{\frac{3\pi}{2}} \!\int\limits_{\psi_i(\frac{\sqrt{3}R}{2})}^{\sqrt{3}R} \!\frac{\mathbbm{1}_{\{i<\frac{\sqrt{3}R}{\Delta y}-\frac{1}{2}\}} \; 2 \; C(\gamma,i)   d\gamma  d\theta_2 }{1+\gamma^{\eta}(s P G(\theta_1)G(\theta_2))^{-1}}  +\int\limits_{0}^{2\pi} \! \int\limits_{\psi_i(\sqrt{3}R)}^{\infty} \!\!\!\! \frac{ C(\gamma,i)   d\gamma  d\theta_2}{1+\gamma^{\eta} (s P G(\theta_1)G(\theta_2))^{-1}} \right) d\theta_1\right\}
\end{aligned}
\end{equation}
\normalsize
\hrule
\vspace{1mm}
\end{figure*}
\small
\begin{equation}\label{comp} 
C(\gamma,i)=\frac{1}{\sqrt{\gamma^2+2\sqrt{\gamma^2-(i \Delta y+\frac{\Delta y}{2})^2}+1}-\gamma}.
\end{equation}
\normalsize
\begin{IEEEproof}
See Appendix A.
\end{IEEEproof}
\end{lemma}

Utilizing the LT in Lemma~\ref{lemm1}, the segment transmission success probability is characterized in the following theorem.
\begin{theorem}
For the D-GW and D-N scenario with constant transmission power $(P)$ and transmission rate $\mathcal{R}_m$, the segment transmission success probability with the 1D-PPPs approximation can be expressed as in \eqref{ee22} presented at the top of next page.
\begin{figure*}
\small
\begin{equation}\label{ee22}
\begin{aligned}[b]
\!\!\!\!\!\!\!p_m\!=\exp\!\left\{-\frac{\Xi_m \sigma^2}{4 P r^{-\eta}_o}\!\!- \sum\limits_{i=0}^{\infty} \int\limits_{0}^{2\pi} \!\! \frac{1}{\pi^2 \Delta x  \mathcal{N}_G} \! \left(   \int\limits_{\frac{\pi}{2}}^{\frac{3\pi}{2}} \!\!\int\limits_{\psi_i(\frac{\sqrt{3}R}{2})}^{\sqrt{3}R} \!\!\!\!\! \frac{\mathbbm{1}_{\{i<\frac{\sqrt{3}R}{\Delta y}-\frac{1}{2}\}} \; 2 C(\gamma,i)   d\gamma  d\theta_2 }{1+4 \gamma^{\eta}( \Xi_m r_o^{\eta} G(\theta_2)G(\theta_1))^{-1}}  \!+\!\int\limits_{0}^{2\pi}\!\! \int\limits_{\psi_i(\sqrt{3}R)}^{\infty} \!\!\!\!\!\! \frac{ C(\gamma,i)   d\gamma  d\theta_2}{1+4 \gamma^{\eta} ( \Xi_m r_o^{\eta} G(\theta_2)G(\theta_1))^{-1}}\right) \! d\theta_1\!\right\}
\end{aligned}
\end{equation}
\normalsize
\hrule
\end{figure*}

\begin{IEEEproof}
Due to the exponential distribution of $h_o$ in \eqref{third33}, the segment transmission success probability in \eqref{third} can be expressed as $p_m\!=\!\exp\!\left\{\frac{-\Xi_m \sigma^2}{4 P r^{-\eta}_o}\right\}\! \times \mathcal{L}_{I_{agg}}\left(\frac{\Xi_m r^\eta_\circ}{4 P}\right)$, where $\mathcal{L}_{I_{agg}}\left(\cdot\right)$ is the LT of the aggregate interference given in Lemma~\ref{lemm1}.
\end{IEEEproof}
\end{theorem}

The segment success probability in \eqref{ee22} requires directive antennas at the gateways and devices. However, directive antennas may not be feasible to be implemented in low-cost and small-size devices. In this case, the segment success probability is characterized in the following corollary. 
\begin{corollary}
For the D-GW and O-N scenario with constant transmission power $(P)$ and transmission rate $\mathcal{R}_m$, the segment transmission success probability for the 1D-PPPs approximation can be expressed as
\begin{small}
\begin{equation}\label{ee3}
\begin{aligned}[b]
\!\!p_m\!=\!\exp\!\left\{\!-\frac{\Xi_m \sigma^2}{2 P r^{-\eta}_o}\! - \!\!\sum\limits_{i=0}^{\infty}\frac{2}{ \pi \Delta x  \mathcal{N}_G}\!\!\! \int\limits_{\psi_i(\frac{\sqrt{3}R}{2})}^{\infty} \! \int\limits_{0}^{2\pi} \!\! \frac{ C(\gamma,i)    d\theta d\gamma}{1+2 \gamma^{\eta} ( \Xi_mr_o^{\eta} G(\theta))^{-1}}\!  \right\}
\end{aligned}
\end{equation}
\end{small}
\begin{IEEEproof}
The SINR for the D-GW and O-N scenario can be obtained from \eqref{third33} by replacing the Tx antenna gains with unity for the intended and interfering links. Following the same steps as in Lemma~1 and Theorem~1, the expression in~\eqref{ee3} can be obtained. 
\end{IEEEproof}
\end{corollary}

The least-cost network deployment can be attained via omni-directional gateways and devices. In such scenario, the segment transmission success probability is characterized in the following corollary. 
\begin{corollary}
For the O-GW and O-N scenario with constant transmission power $(P)$ and transmission rate $\mathcal{R}_m$, the segment transmission success probability for the 1D-PPPs approximation can be expressed as
\begin{small}
\begin{equation}\label{ee4}
\begin{aligned}[b]
p_m\!=\!\exp\!\left\{-\frac{\Xi_m \sigma^2}{P r^{-\eta}_o}-\sum\limits_{i=0}^{\infty} \frac{4}{ \Delta x  \mathcal{N}_G} \!\!\!\! \int\limits_{\psi_i(\frac{\sqrt{3}R}{2})}^{\infty}  \!\!\!\! \frac{C(\gamma,i)    }{1+\gamma^{\eta} (\Xi_m r_o^{\eta})^{-1}}  d\gamma \right\}
\end{aligned}
\end{equation}
\end{small}
\begin{IEEEproof}
The SINR for the O-GW and O-N scenario can be obtained from \eqref{third33} by replacing the Tx and Rx antenna gains with unity for the intended and interfering links. Following the same steps as in Lemma~1 and Theorem~1, the expression in~\eqref{ee4} can be obtained. 
\end{IEEEproof}
\end{corollary}

\begin{remark}
The 1D-PPPs approximation enables tractable characterization for the segment transmission success probabilities for different network scenarios in Theorem~1 and Corollaries~1~and~2. However, all of the obtained expressions are of high numerical complexity because of the exponential expression that involves infinite summation of infinite integral functions. To further facilitate the analysis and reduce the involved numerical complexity, the 2D-PPP approximation is presented in the next section. 
\end{remark}

\subsection{2D-PPP Model with Constant Tx Power}

This section utilizes Approximations~\ref{app_orint} and \ref{app_2DPPP} to characterize the success probability in \eqref{third} with constant transmission power scheme. Exploiting the 2D-PPP approximation, the LT of the aggregate interference seen at the test gateway is characterized in the following lemma.
\begin{lemma} \label{lemm2}
For the D-GW and D-N scenario with constant transmission power $(P)$, the LT of the 2D-PPP aggregate interference observed at the test gateway located at the origin can be expressed as \eqref{lap_area_2} presented at the top of next page, where ${}_2F_1(\cdot)$ is the Gauss hypergeometric function.
\begin{figure*}
\small
 \begin{equation}\label{lap_area_2}
  \begin{aligned}[b]
\!\!\!\mathcal{L}_{Iagg} (s)\!=\!&\exp\left\{\!-\!\int\limits_{0}^{2\pi} \!\!\! \frac{ s P G(\theta_1)\; (\sqrt{3} R)^{2-\eta}}{4 \pi (\eta -2) \Delta x \Delta y  \mathcal{N}_G }\left[ \int\limits_{0}^{2\pi} \!\! 2 G(\theta_2)  \; {_2F_1} 
\Big(1,1-\frac{2}{\eta};2-\frac{2}{\eta};\frac{-sP G(\theta_2)G(\theta_1) }{(\sqrt{3}R)^{\eta}}\Big) d\theta_2 + \right. \right.  \\&
\quad \quad \left. \left. \left. \int\limits_{\pi/2}^{\frac{3 \pi}{2}} \!\! G(\theta_2) \left( 2^\eta \;  {_2F_1}\Big(1,1-\frac{2}{\eta};2-\frac{2}{\eta};\frac{-sP G(\theta_2)G(\theta_1) }{(\frac{\sqrt{3}R}{2})^{\eta}}\Big)- 4 \;  {_2F_1}
\Big(1,1-\frac{2}{\eta};2-\frac{2}{\eta};\frac{-sP G(\theta_2)G(\theta_1) }{(\sqrt{3}R)^{\eta}}\Big) \right) d\theta_2 \right] \!d\theta_1\!\! \right.\right\}
  \end{aligned}
\end{equation}
\hrule
\normalsize
\end{figure*}

\begin{IEEEproof}
See Appendix B.
\end{IEEEproof}
\end{lemma}
Utilizing the LT in Lemma~\ref{lemm2}, the segment transmission success probability is characterized in the following theorem.
\begin{theorem}
For the D-GW and D-N scenario with constant transmission power $(P)$ and transmission rate $\mathcal{R}_m$, the segment transmission success probability with the 2D-PPP approximation can be expressed as in \eqref{sp_area}  presented at the top of next page.
\begin{figure*}
\small
\begin{equation}\label{sp_area}
  \begin{aligned}[b]
\!\!\!p_m\!=\exp&\!\left\{-\frac{\Xi_m r^{\eta}_o \sigma^2}{4 P }\!-\!\int\limits_{0}^{2\pi} \!\!\! \frac{ \Xi_m r_o^\eta  G(\theta_1)\; (\sqrt{3} R)^{2-\eta}}{16 \pi (\eta -2) \Delta x \Delta y  \mathcal{N}_G }\left[ \int\limits_{0}^{2\pi} \!\! 2 G(\theta_2)  \; {_2F_1} 
\Big(1,1-\frac{2}{\eta};2-\frac{2}{\eta};\frac{-\Xi_m r_o^\eta  G(\theta_2)G(\theta_1) }{ 4 (\sqrt{3}R)^{\eta}}\Big) d\theta_2 + \right. \right.  \\&
\quad \quad \left. \left. \left. \int\limits_{\pi/2}^{\frac{3 \pi}{2}} \!\! G(\theta_2) \left( 2^\eta \;  {_2F_1}\Big(1,1-\frac{2}{\eta};2-\frac{2}{\eta};\frac{-\Xi_m r_o^\eta  G(\theta_2)G(\theta_1) }{ 4 (\frac{\sqrt{3}R}{2})^{\eta}}\Big)- 4 \;  {_2F_1}
\Big(1,1-\frac{2}{\eta};2-\frac{2}{\eta};\frac{-\Xi_m r_o^\eta  G(\theta_2)G(\theta_1) }{4 (\sqrt{3}R)^{\eta}}\Big) \right) d\theta_2 \right] \!d\theta_1\!\! \right.\right\}
  \end{aligned}
\end{equation}
\normalsize
\hrule
\end{figure*}
\begin{IEEEproof}
The theorem is proved in a similar way to Theorem 1 but by utilizing the LT of the 2D-PPP aggregate interference given in Lemma 2. 
\end{IEEEproof}
\end{theorem}

For cost reduction, the two scenarios of D-GW with O-N and O-GW with O-N are given in the following two corollaries. 

\begin{corollary}
For the D-GW and O-N scenario with constant transmission power $(P)$ and transmission rate $\mathcal{R}_m$, the segment transmission success probability for the 2D-PPP approximation can be expressed as
\small
\begin{equation}\label{cov_area1}
 \begin{aligned}[b]
\!\! \!\! \!\!p_m=&\exp\left\{-\frac{\Xi_m r^{\eta}_o \sigma^2}{2 P} -\int_{0}^{2\pi} \frac{\Xi_m  r_o^{\eta}G(\theta)(\frac{\sqrt{3}R}{2})^{2-\eta}}{2 \Delta x \Delta y  \mathcal{N}_G  (\eta -2)}  \right.\\
& 
\qquad\qquad\qquad \left.{_2F_1}\left(1,1-\frac{2}{\eta};2-\frac{2}{\eta};\frac{- \Xi_m r_o^{\eta} G(\theta) }{2(\frac{ \sqrt{3}R}{2})^{\eta}}\right)d\theta \right\}
 \end{aligned}
\end{equation}
\normalsize
\begin{IEEEproof}
The corollary is proved in a similar way to Corollary~1 but by following the steps of Lemma~2 and Theorem~2. 
\end{IEEEproof}
\end{corollary}
\begin{corollary}
For the O-GW and O-N scenario with constant transmission power $(P)$ and transmission rate $\mathcal{R}_m$, the segment transmission success probability for the 2D-PPP approximation can be expressed as
\small
\begin{equation}\label{covr_area_omni2}
\begin{aligned}[b]
\!\!\!\!\!p_m=&\exp\left\{-\frac{\Xi_m r^{\eta}_o \sigma^2}{ P}-\frac{2\pi \Xi_m r_o^{\eta}(\frac{\sqrt{3}R}{2})^{2-\eta}}{\Delta x \Delta y  \mathcal{N}_G (\eta -2)} \right. \\
& \qquad\qquad\qquad \qquad \qquad  \left. {_2F_1}
\left(1,1-\frac{2}{\eta};2-\frac{2}{\eta};\frac{-\Xi_m r_o^{\eta}}{(\frac{\sqrt{3}R}{2})^{\eta}}\right)\right\}
\end{aligned}
\end{equation}
\normalsize
\begin{IEEEproof}
The corollary is proved in a similar way to Corollary~2 but by following the steps of Lemma~2 and Theorem~2. 
\end{IEEEproof}
\end{corollary}

\begin{remark}
The expressions in Theorem 2 and Corollary 3 only involve finite integrals. A closed-form expression is obtained for the O-GW and O-N scenario in Corollary 4. Hence, the 2D-PPP approximation provides much simpler and numerically less complex expressions for the segment transmission success probability when compared to the 1D-PPPs approximation. 
\end{remark}

\begin{remark}
{It is worth highlighting that the two proposed approximate models 1D-PPP and 2D-PPP utilize stochastic geometry analysis for tractability. Hence, the developed approximate models are applicable to networks with random deployment as done [7]-[9], [14], [15], [17], [20], [23]. However, this paper shows that such tractable stochastic geometry analysis can also be utilized for regularly deployed networks if careful consideration are given to the proximate interfering devices, which dominate the aggregate interference.}
\end{remark}

\subsection{The Effect of Power Control}\label{power_control_sec}

The segment transmission success probabilities given in Theorems 1 and 2 as well as Corollaries 1 to 4 are all functions of the device distance from its serving gateway. Due to the constant transmission power, devices that are closer to their gateways enjoy better success probability than those which are farther away from their gateways. To unify the performance of all devices (i.e., improve fairness), path-loss inversion power control is adopted. Such power control enables all devices to maintain a unified target average received power ($\rho$) at their serving gateways, and hence, the SINR in \eqref{third33} can be re-written as 
\begin{equation}\label{sinr_pc}
\text{SINR}=\frac{\rho h_o G(\theta_{T_\circ}) G(\theta_{R_\circ})} {\sigma^2+\sum\limits_{v_i \in \Psi_{2D}} \mathbbm{1}_{\{\text{is active}\}}  P_i h_{i} \left\|v_{i}\right\|^{-\eta} G(\theta_{T_{i}}) G(\theta_{R_{i}})}.
\end{equation}
Note that power control affects the aggregate interference as $P_i$ in \eqref{sinr_pc} becomes a random variable that depends on the distance between the $i^{th}$ interfering device and its serving gateway. The LT of the aggregate interference term in \eqref{sinr_pc} is characterized in the following lemma.\footnote{Owing to its simplicity, only the 2D-PPP approximation is unitized to study the effect of power control.}
 \begin{lemma} \label{lemm3}
For the D-GW and D-N scenario with path-loss inversion power control, the LT of the 2D-PPPs aggregate interference observed at the test gateway located at the origin can be expressed as \eqref{lap_pc} presented at the top of next page, where $\mathbb{E}\{r^2\}$ is the second moment of the distances between the active devices and their serving gateways. Due to the uniformly randomized scheduling and network geometrical symmetry, we have
\begin{equation*}
    \mathbb{E}\{r^2\} = \frac{2}{ \mathcal{N}_G} \sum_{i=1}^Y \sum_{j} \mathbbm{1}_{\{ v_{ij} \in \Psi_i \cap \; \tinyvarhexagon_\circ\}} \left\|v_{ij} \right\|^2,
\end{equation*}
where $\tinyvarhexagon_\circ$ denote the coverage hexagonal region of the test gateway.
\begin{figure*}
\small
 \begin{equation}\label{lap_pc}
   \begin{aligned}[b]
\!\!\!\mathcal{L}_{Iagg} (s)\!=\!&\exp\left\{\!-\!\frac{\Xi_m \sigma^2}{4\rho}\!-\!\int\limits_{0}^{2\pi} \frac{s \rho \mathbb{E}\left\{r^2\right\}  G(\theta_1)}{ \pi (\eta -2) \Delta x \Delta y  \mathcal{N}_G }     \left[ \int\limits_{0}^{2\pi} \frac{3^{2-\eta} G(\theta_2)}{2} \; {_2F_1}
\Big(1,1-\frac{2}{\eta};2-\frac{2}{\eta};\frac{-s \rho G(\theta_2)G(\theta_1) }{ 3^{\eta}}\Big) d\theta_2 +  \right. \right.\\ 
& \qquad \qquad \qquad  \left. \left. 
\int\limits_{\frac{\pi}{2}}^{\frac{3 \pi}{2}}  \!\! G(\theta_2)\left( {_2F_1}\Big(1,1-\frac{2}{\eta};2-\frac{2}{\eta};- s \rho G(\theta_2)G(\theta_1)\Big) -  3^{2-\eta} {_2F_1}
\Big(1,1-\frac{2}{\eta};2-\frac{2}{\eta};\frac{-s \rho G(\theta_2)G(\theta_1) }{ 3^{\eta}}\Big) \right) d\theta_2 \right]d\theta_1\! \right\}
  \end{aligned}
\end{equation}
\normalsize
\hrule
 \end{figure*}

 \begin{IEEEproof}
 See Appendix C.
 \end{IEEEproof}
\end{lemma}

Utilizing the LT in Lemma~\ref{lemm3}, the segment transmission success probability is characterized in the following theorem.
\begin{theorem}
For the D-GW and D-N scenario with path-loss inversion power control and transmission rate $\mathcal{R}_m$, the segment transmission success probability with the 2D-PPPs approximation can be expressed as in \eqref{sp_pc} presented at the top of next page.
\begin{figure*}
\small
 \begin{equation}\label{sp_pc}
  \begin{aligned}[b]
\!\!\!\!\!\!p_m\!=\!&\exp\left\{\!-\frac{\Xi_m \sigma^2}{4\rho}\!-\!\int\limits_{0}^{2\pi} \frac{\Xi_m \mathbb{E}\left\{r^2\right\}  G(\theta_1)}{4 \pi (\eta -2) \Delta x \Delta y  \mathcal{N}_G }     \left[ \int\limits_{0}^{2\pi} \frac{3^{2-\eta} G(\theta_2)}{2} \; {_2F_1}
\Big(1,1-\frac{2}{\eta};2-\frac{2}{\eta};\frac{-\Xi_m G(\theta_2)G(\theta_1) }{4 \times 3^{\eta}}\Big) d\theta_2 +  \right. \right.\\ 
& \qquad \qquad \qquad  \left. \left. 
\int\limits_{\frac{\pi}{2}}^{\frac{3 \pi}{2}}  \!\! G(\theta_2)\left( {_2F_1}\Big(1,1-\frac{2}{\eta};2-\frac{2}{\eta};\frac{- \Xi_m G(\theta_2)G(\theta_1)}{4}\Big) -  3^{2-\eta} {_2F_1}
\Big(1,1-\frac{2}{\eta};2-\frac{2}{\eta};\frac{-\Xi_m  G(\theta_2)G(\theta_1) }{4 \times 3^{\eta}}\Big) \right) d\theta_2 \right]d\theta_1\! \right\}
  \end{aligned}
\end{equation}
\normalsize
\hrule
 \end{figure*}

\begin{IEEEproof}
The theorem is proved in a similar way to Theorem 1 but by utilizing the LT of the 2D-PPP aggregate interference with power control given in Lemma 3. 
\end{IEEEproof}
\end{theorem}

Following the same approach as in Corollaries 1 to 4, the segment transmission success probability for the special cases of D-GW with O-N and O-GW with O-N are given in the following two corollaries.

\begin{corollary}
For the D-GW and O-N scenario with path loss inversion power control and transmission rate $\mathcal{R}_m$, the segment transmission success probability for the 2D-PPP approximation can be expressed as
 \begin{equation}\label{sp1_pc22}
 \small
  \begin{aligned}[b]
\!\!\!\!\!\!\! p_m=&\exp\Bigg\{- \frac{\Xi_m \sigma^2}{2\rho} - \frac{\Xi_m \mathbb{E}\{r^2\}}{2 (\eta-2) \Delta x \Delta y  \mathcal{N}_G}  \int\limits_{0}^{2\pi}  G(\theta) \times \\
& \qquad \qquad \qquad \quad
{_2F_1}
\Big(1,1-\frac{2}{\eta};2-\frac{2}{\eta};\frac{- \Xi_m G(\theta)  }{2} \Big)d\theta \Bigg\}
  \end{aligned}
\end{equation}
\normalsize
\begin{IEEEproof}
The corollary is proved in a similar way to Corollary~1 but by following the steps of Lemma~3 and Theorem~3. 
\end{IEEEproof}
\end{corollary}

\begin{corollary}
For the O-GW and O-N scenario with path loss inversion power control and transmission rate $\mathcal{R}_m$, the segment transmission success probability for the 2D-PPP approximation can be expressed as
 \begin{equation}\label{sp1_pc33}
 \small
  \begin{aligned}[b]
p_m\!=\!&\exp\Bigg\{\!-\! \frac{\Xi_m \sigma^2}{\rho}\! -\! \frac{2 \pi \Xi_m \mathbb{E}\{r^2\} \; {_2F_1}
\Big(1,1-\frac{2}{\eta};2-\frac{2}{\eta}; - \Xi_m \Big)}{ (\eta-2) \Delta x \Delta y  \mathcal{N}_G} 
 \Bigg\}
  \end{aligned}
\end{equation}
\normalsize
\begin{IEEEproof}
The corollary is proved in a similar way to Corollary~2 but by following the steps of Lemma~3 and Theorem~3. 
\end{IEEEproof}
\end{corollary}

\section{ Tx Rate \& Delay}\label{qm}

Treating interference as noise, the maximum achievable Tx rate is defined by the ergodic capacity $\mathcal{C} = \zeta \times W\times \mathbb{E}\{\log_2(1+\text{SINR})\}$ bits/sec. To achieve $\mathcal{C}$ bits/sec, the transmitting devices require instantaneous knowledge (i.e., every time slot) of the SINR realization in order to adapt their Tx rate and alleviate outages~\cite{ER_Haenggi}. Indeed this is infeasible for large-scale IoT networks. Due to the absence of instantaneous SINR feedback, the IoT devices operate at a fixed transmission rate $\mathcal{R}_m < \mathcal{C} $ that is subject to outages (i.e., with probability $[1-p_m]$), and hence, the uplink link throughput is expressed as 
\begin{equation}\label{ER11}
  \begin{aligned}
\mathcal{T}_m &= \mathbb{P}\{\text{SINR}> \Xi_m\} \times \zeta \; W  \; \log_2(1+\Xi_m) = p_m \times \mathcal{R}_m 
  \end{aligned}
\end{equation}
where $p_m=\mathbb{P}\{\text{SINR}> \Xi_m\}$ is calculated in Section~\ref{sp_m} for the different network scenarios and $\mathcal{R}_m$ is given in \eqref{rate}.  The expressions in \eqref{rate}  and  \eqref{ER11} advocate that the Tx rate $\mathcal{R}_m$ should be selected in light of the data granularity parameters $L$ and $T_r$ as well as the SINR dependent Tx success probability $p_m$.

To account for temporal generation and departure of data packets, we utilize the Matrix-Analytic Method (MAM) (see \cite{alfa2016applied}) to develop a novel discrete-time PH/PH/1 queueing system that tracks the packet generation, rate-dependent segmentation, device scheduling, and SINR-aware segment departure from the queue of each device. From the device perspective, the temporal resolution at which a change can occur in the queue is determined by the transmission cycle $T_c = \mathcal{N}_G T_s$ seconds. Recall that, as shown in Fig.~\ref{scheduling_cycle}, periodic arrivals occur every $T_r = T_a \times \mathcal{N}_G T_s = T_a \times T_c$. Furthermore, each device is granted a single transmission attempt within a transmission cycle, which allows no more than one segment departure every $T_c$. Consequently, the time resolution of the developed discrete-time PH/PH/1 queueing model is $T_c = \mathcal{N}_G T_s$.

The periodic generation of packets every $T_a$ transmission cycles is modeled via a PH type distribution with initialization vector $\boldsymbol{\alpha} = [1 \;\; 0 \;\; 0\; \cdots  \;0]$ of size $T_a$, transient transition matrix $\mathbf{K}$ of size $T_a \times T_a$ and an absorption vector $\mathbf{k}$ of size $T_a \times 1$, which are given by
\begin{equation}\label{T_mat}
\mathbf{k}=\begin{bmatrix}
0 \\
0 \\
\vdots \\
0  \\
1
\end{bmatrix} \qquad \text{and} \qquad \mathbf{K}=\begin{bmatrix}
0 & 1 & 0 &\cdots & 0\\
0 & 0 & 1 &\cdots & 0\\
\vdots  & \vdots  & \ddots  & \ddots & \vdots\\
0  & 0  & \cdots  & 0 & 1\\
0 & 0 & 0 & 0 & 0
\end{bmatrix}.
\end{equation}

\begin{remark}
The transient matrix $\mathbf{K}$ tracks the deterministic time evolution between two consecutive packets generation.  Hence, only upper off-diagonal ones are the non-zero elements in $\mathbf{K}$. After recording the evolution of $T_a$ transmission cycles with $\mathbf{K}$, an absorption occurs with $\mathbf{k}$ to denote a new packet arrival. 
\end{remark}

A packet departure of size $L$ at rate $\mathcal{R}_m$ is modeled via another PH type distribution with initialization vector $\boldsymbol{\beta}_m = [1 \;\; 0 \;\; 0\; \cdots  \;0]$ of size $m$, a transient transition matrix $\mathbf{S}_{m}$ of size $m \times m$, an absorption vector $\mathbf{s_m}$ of size $m \times 1$, which are given by
\begin{equation}\label{sn}
\mathbf{s}_m=\begin{bmatrix}
0 \\
0 \\
\vdots \\
0  \\
p_m
\end{bmatrix} \quad \text{and} \quad
\mathbf{S}_{m}=\begin{bmatrix}
\overline{p}_{m} & p_{m} & 0 &\ldots & 0\\
0 &\overline{p}_{m}  &  p_{m} &\ddots & 0\\
\vdots  & \ddots  & \ddots  & \ddots & \vdots\\
0  & \ldots  & 0  & \overline{p}_{m}  & p_{m} \\
0 & \ldots & 0 & 0 & \overline{p}_{m}
\end{bmatrix},
\end{equation}
where $\bar{p}_m=1-p_m$ and $p_m$ is the segment transmission success probability calculated in Section~\ref{sp_m} for the different network scenarios.

\begin{remark}
At rate $\mathcal{R}_m$, a packet is partitioned into $m$ segments that need to be all successfully delivered. The transient matrix $\mathbf{S}_m$ tracks the probabilistic segment departure of each of the $m$ segments from the devices queue. The diagonal elements of $\bar{p}_m$ denote transmission failure probability, and hence, no progress in segment departure is recorded. On the other hand, the off-diagonal elements are $p_m$ to record the progress due to a single segment's successful departure. After recording the successful departure of $m$ segments, the absorption vector $\mathbf{s}_m$ implies the successful departure of the entire packet. Note that $\mathbf{S}_m$ and $\mathbf{s}_m$ define a rate-sensitive and SINR-aware departure process that accounts for the network parameters (e.g., packet size, antenna directivity, power control, and a number of devices per gateway).   
\end{remark}

Using the MAM \cite{alfa2016applied}, we construct the PH/PH/1 queueing model with state-space $(q,d,t)$, where $q \in \mathbb{N}$ is the number of packets in the devices buffer, $d \in \{1,\;2,\; \cdots,\; m\}$ is the number left-over segments that belongs to the packet in transmission, and $t \in \{0,\; 1,\;  \cdots,\; T_a\}$ is the number of time epochs (i.e., transmission cycles) that elapsed since the last packet arrival. Since only one packet arrival and/or departure can occur in a single time epoch, the constructed PH/PH/1 queueing model is a quasi-birth-death (QBD) process with the following transition matrix  
\begin{equation}\label{qbd}
\mathbf{P}_m=\begin{bmatrix}
\mathbf{B} & \mathbf{C} & \mathbf{0} & \mathbf{0} & \mathbf{0} & \ldots\\
\mathbf{E} & \mathbf{A_1} & \mathbf{A_0} & \mathbf{0} &\mathbf{ 0}& \ddots\\
\mathbf{0} & \mathbf{A_2} & \mathbf{A_1} & \mathbf{A_0} & \mathbf{0} & \ddots\\
\vdots & \ddots & \ddots & \ddots & \ddots& \ddots
\end{bmatrix}, 
\end{equation}
where the sub-matrices of $\mathbf{P}_m$ are defined as follows
\begin{align*}
\mathbf{B}\!=\! \mathbf{K}, \quad  \mathbf{E}\!=\!\mathbf{K} \otimes \mathbf{s}_{m}, \quad 
\mathbf{C} \!=(\mathbf{k} \boldsymbol{\alpha})\otimes \boldsymbol{\beta}_{m}, \quad \mathbf{A_0} \!=\! (\mathbf{k} \boldsymbol{\alpha}) \otimes \mathbf{S}_{m}  \\
\mathbf{A_1} \!=\!(\mathbf{k} \boldsymbol{\alpha}) \otimes (\mathbf{s}_{m} \boldsymbol{\beta}_{m}) + \mathbf{K} \otimes \mathbf{S}_{m},\quad \text{and} \quad
\mathbf{A_2} \!=\! \mathbf{K}  \otimes (\mathbf{s}_{m} \boldsymbol{\beta}_{m}) 
\end{align*}
and $\otimes$ is the Kronecker product operator. In particular, the $T_a \times T_a$ sized $\mathbf{B}$ matrix, the $m T_a \times T_a$ sized  $\mathbf{E}$ matrix, and $T_a \times m T_a$ sized $\mathbf{C}$ martix are the sub-stochastic boundary matrices that track, respectively,
the transitions among the states in $t$ for $q=0$, the transition from $q=0$ to $q=1$ at $t=T_a$, and the transition from $q=1$ to $q=0$ at the departure of the last segment, which can occur at any phase $t$. The matrices $\mathbf{A_0}$, $\mathbf{A_1}$, and $\mathbf{A_2}$ are all of the same size $m T_a \times m T_a$ to track, respectively, the upward transitions from $q$ to $q+1$, the transitions within the same level $q$, and the downward transitions from $q+1$ to $q$. Note that the matrices $\mathbf{A_0}$, $\mathbf{A_1}$, and $\mathbf{A_2}$ encompass all transitions that track segment departures $d$ as well as the time evolution $t$ from the last packet arrival. 

The transition matrix $\mathbf{P}_m$ can be constructed for each transmission rate $\mathcal{R}_m$ within each network scenario, namely, the D-GW with D-N, the D-GW with O-N, and the O-GW with O-N. For the constant power control scheme, each device location would have a different transition matrix $\mathbf{P}_m$ due to the location-dependent transmission success probability $p_m$. However, for the power control scheme, all the devices would have the same transition matrix $\mathbf{P}_m$ due to the location-independent transmission success probability $p_m$ enforced by the path-loss inversion power control. 

Before conducting the steady-state analysis, the stability of the queueing model has to be checked. That is, it has to be ensured that the queue utilization is less than unity. Otherwise, the queue size $q$ will grow indefinitely and lead to infinite delay. The queueing system stability is given in the following lemma 

\begin{lemma} \label{stability_lem}
The queue utilization for the periodic arrival PH type process $(\boldsymbol{\alpha},\mathbf{K})$ defined in \eqref{T_mat} and the rate-aware and SINR-sensitive departure PH type process $(\boldsymbol{\beta},\mathbf{S}_m)$ defined in \eqref{sn} can be expressed as
\begin{equation}
\rho_m = \frac{m}{p_m \times T_a} =  \frac{m  \mathcal{N}_G T_s}{p_m T_r} .
\end{equation}
The queueing model at rate $\mathcal{R}_m$ is stable if and only if $\rho_m<1$.
\begin{IEEEproof}
Queue utilization is defined as the ratio between the mean arrival and mean departure rates. For the arrival PH type process, the mean arrival rate is given by the reciprocal of the meantime to absorption. Similarly, the average departure rate is given by the reciprocal of the meantime to absorption. The mean time to absorption for the arrival and departure processes are given by, respectively, $\boldsymbol{\alpha} \left(\mathbf{I} - \mathbf{K}\right)^{-1} \mathbf{1}$ and $\boldsymbol{\beta} \left(\mathbf{I} - \mathbf{S}_m\right)^{-1} \mathbf{1}$, where $\mathbf{I}$ is the identity matrix and $\mathbf{1}$ is the column vector of ones with the proper sizes. After some mathematical manipulation and accounting for the special structure of $\boldsymbol{\alpha}$, $\boldsymbol{\beta}$, $\mathbf{K}$ and $\mathbf{S}_m$ the lemma is proved. 
\end{IEEEproof}
\end{lemma}

For stable queues (i.e.,  $\rho_m<1$), the steady-state queue distribution can be obtained by solving the following set of linear equations 
\begin{align} \label{ss_queue}
\boldsymbol{\pi} =\boldsymbol{\pi} \mathbf{P}_m \quad \text{and} \quad \boldsymbol{\pi} \mathbf{1} = 1,
\end{align}
where $\boldsymbol{\pi}=[\boldsymbol{\pi}_0 \; \boldsymbol{\pi}_1 \; \boldsymbol{\pi}_2 \cdots \boldsymbol{\pi}_q \cdots]$ encompass the joint distribution of three variables $(q,d,t)$. In particular, $\boldsymbol{\pi}_0$ is of length $1 \times T_a $ to capture all the probabilities of the phases $t$ when there are no packets in the queue. For $q\geq1$, the vector $\boldsymbol{\pi}_q$ is of length $1 \times m T_a$ to capture all the probabilities of the number of left-over segments ($d$) of the packet in transmission and the elapsed time epochs $t$ from the last packet arrival when there are $q$ packets in the system. Since $\boldsymbol{\pi}$ is of infinite size, the system in \eqref{ss_queue} is solved via MAM \footnote{ {Note that MAM computation is conducted offline to characterize the network
performance and come-up with long-term network design (e.g., $m$, $T_a$, $T_r$  and $L$). Such network design do not need to be changed as long as the underlying network parameters (e.g, fading distribution, gateways density and devices density) remain fixed.}} as shown in the following theorem

\begin{theorem}
 Let $\mathbf{R}$ be MAM rate matrix defined as the minimal non-negative solution of $\mathbf{R} =\mathbf{A_0}+\mathbf{R} \mathbf{A_1}+\mathbf{R}^2 \mathbf{A_2}$, then $\boldsymbol{\pi}$ in \eqref{ss_queue} is given by 
	 	\begin{equation}
  	 	   \label{SS_sol_theorem}
  	 	\!\!\!\!\!\!\!\!\!\!\!\!	\boldsymbol{\pi}_{q} \! =\!\left\{\begin{matrix}
  	 	\boldsymbol{\pi}_{1} \mathbf{E} (\mathbf{I}-\mathbf{B})^{-1} & \text{for} \; q=0 \\ 
  	 	&\\
  	 	\text{Nul}\Big(\mathbf{Q} \Big)& \text{for}\; q=1 \\
  	 	&\\
  	 	\boldsymbol{\pi}_{1} \mathbf{R}^{q-1} &  \text{for}\; q\geq 2
  	 	\end{matrix}\right., 	
  	 	\end{equation}
where $\text{Nul}(\mathbf{Q})$ is the null space vector of the matrix $\mathbf{Q}=\left(\mathbf{E} (\mathbf{I}-\mathbf{B})^{-1} \mathbf{C}+ \mathbf{A_1}+\mathbf{R}\mathbf{A_2}\right)^\text{T}-\mathbf{I}$ that satisfies $\boldsymbol{\pi}_0 \mathbf{1}+\boldsymbol{\pi}_1(\mathbf{I}-\mathbf{R})^{-1}\mathbf{1} = 1$. 
\begin{IEEEproof}
By definition of the rate matrix \cite{alfa2016applied}, we have $\boldsymbol{\pi}_q = \boldsymbol{\pi}_1 \mathbf{R}^{q-1}$ for all $q>1$. Substituting  $\boldsymbol{\pi}_2 = \boldsymbol{\pi}_1 \mathbf{R}$, the boundary vectors of $\boldsymbol{\pi} =\boldsymbol{\pi} \mathbf{P}_m$ can be written as 
\begin{equation}
    [\boldsymbol{\pi}_0 \;\; \boldsymbol{\pi}_1]=[\boldsymbol{\pi}_0 \;\; \boldsymbol{\pi}_1]\begin{bmatrix}
\mathbf{B} & \mathbf{C}  \\
\mathbf{E} & \mathbf{A_1} + \mathbf{R} \mathbf{A_2}
\end{bmatrix},
\end{equation}
which are solved to obtain $\boldsymbol{\pi}_0$ and $\boldsymbol{\pi}_1$ in \eqref{SS_sol_theorem}. Furthermore, the normalization condition $\boldsymbol{\pi} \mathbf{1} = 1$ implies that $\sum_i \boldsymbol{\pi}_i \mathbf{1} = \boldsymbol{\pi}_0 \mathbf{1}+ \sum _{i=1}^\infty \boldsymbol{\pi}_1 \mathbf{R}^{i-1} \mathbf{1}= \boldsymbol{\pi}_0 \mathbf{1}+\boldsymbol{\pi}_1(\mathbf{I}-\mathbf{R})^{-1}\mathbf{1} = 1$ which has to be satisfied by the solution. 
\end{IEEEproof}
\end{theorem}
The complicated nature of $\mathbf{A_0}$, $\mathbf{A_1}$, and $\mathbf{A_2}$ is a common problem of the PH/PH/1 queues, and hence, there is no explicit expression for the rate matrix $\mathbf{R}$ in most practical scenarios~\cite{alfa2016applied}. However, there are several efficient methods to numerically compute $\mathbf{R}$ given in Theorem 4 such as the cyclic reduction, the logarithmic reduction, or invariant subspace methods \cite{alfa2016applied}.

The steady-state probability vector $\boldsymbol{\pi}$ provides a full characterization of the queueing model in \eqref{qbd}, which can be utilized to study several key performance indicators. For instance, the average queue size is given by 
\begin{equation} \label{av_size}
    \mu_L=\sum_{i=1}^{\infty}i\boldsymbol{\pi}_i \mathbf{1} = \boldsymbol{\pi}_1 (\mathbf{I}-\mathbf{R})^{-2} \mathbf{1}.
\end{equation}
Leveraging \eqref{av_size} along with Little’s Law,  
the average total delay (i.e., queueing  plus transmission delay) is obtained as \begin{equation}
    \mu_W = T_a \mu_L = T_a \boldsymbol{\pi}_1 (\mathbf{I}-\mathbf{R})^{-2} \mathbf{1}
\end{equation}
 It is worth mentioning that the optimal number of segments ($m^*$) is to be selected such that the total average delay ($\mu_W$) is minimized. Due to the stability condition in Lemma~\ref{stability_lem}, the feasibility range of segmentation is limited to the integer values within the range $1 \leq m \leq \floor{p_m T_a}$. For more detailed delay characterization, $\boldsymbol{\pi}$ can be utilized to derive the full delay distribution of a randomly selected packet in the network. Following \cite[Sec. 5.10]{alfa2016applied}, the probability mass function of the delay can be obtained as 
  \begin{equation}
  \mathbb{P}\{w_q =i \} = \left\{ \begin{matrix}
  \mathbf{y}_0 \mathbf{1}  & \text{for } i=0\\
  & \\
  \sum\limits_{z=1}^{i}\mathbf{y}_z(\mathbf{1}\otimes \mathbf{I}) \mathbf{B}_i^{(z)}\mathbf{1} & \text{for } i\geq 1
  \end{matrix}\right.
\end{equation}
where 
  \begin{equation*}
  \mathbf{y}_z = \left\{ \begin{matrix}
   T_a \Big[\boldsymbol{\pi}_0\mathbf{k} \mathbf{\alpha}+\boldsymbol{\pi}_1(\mathbf{k} \mathbf{\alpha} \otimes \mathbf{s}_m)\Big]  & \text{for } z=0\\
  & \\
  \!\!\!\!\!\! \!\!\!\!\!\!  \!\!\!\!\!\! \!\!\!\!\!\! \!\!\!\!\!\! \boldsymbol{\pi}_i \mathbf{R}^{i-1}\Big(T_a \big[\mathbf{k} \mathbf{\alpha} \otimes \mathbf{S}_m  &  \\
    \qquad \quad \quad \qquad+\mathbf{R} (\mathbf{k} \mathbf{\alpha} \otimes \mathbf{s}_m \mathbf{\beta}_m)\big]\Big)  & \text{for } z\geq 1 
  \end{matrix}\right.
\end{equation*}
and 
	\begin{equation*}
  	 	   \label{B}
  	 	\mathbf{B}_i^{(z)}\! =\!\left\{\begin{matrix}
  	 	\mathbf{S}_m^{i-1}(\mathbf{s}_m \mathbf{\beta}_m) & \text{for} \; z=1 \\ 
  	 	&\\
  	 \mathbf{S}_m \mathbf{B}_{i-1}^{(z)}+(\mathbf{s}_m \mathbf{\beta}_m)\mathbf{B}_{i-1}^{(z-1)} & \text{for}\; 1< z < i \\
  	 	&\\
  	 (\mathbf{s}_m \mathbf{\beta}_m)^z &  \text{for}\; z=i
  	 	\end{matrix}\right..	
  	 	\end{equation*}



\section{Numerical Results \& Simulations }\label{rs}

 \begin{table} 
  \begin{center}
     \caption{ Parameters for numerical demonstration.}
    \begin{tabular}{lcl}
      \hline
      \textbf{Parameter Description} & \textbf{Symbol} & \textbf{Value} \\
      \hline
      {Network size} & {--} & {70 km$^2$} \\
       Number of lines & -- & 30 \\
      Spacing of devices & $\Delta x$ & 25 m \\
      Spacing of lines & $\Delta y$ & 200 m \\
      Hexagonal cell radius & $R$ & 490 m \\
      Path loss exponent & $\eta$ & 4 \\
      Theoretical-to-practical rate &$\zeta$ & 0.8\\
      Nodes per cell & $\mathcal{N}_G$ & 120\\
      Time slot duration & $T_s$ & 10 ms\\
      New packet every $T_a$ cycles  & $T_a$   & 18 Tx cycles\\
      Inter-arrival time & ${T_r}$   & 21.6 s\\
     Bandwidth  &    $W$       &   1 MHz\\
     {Packet size} &   {$L$}       & {80 kbits}\\
    {Average received signal power {with} power control}     & {$\rho$}   & {-100 {dBm}}\\
    {Noise power}    & {$\sigma^2$}   & {-110 {dBm}}\\
       {Constant transmission  power}   & {$P$}  & {1.2 mW}
    \end{tabular}
      \label{table1}
  \end{center}
\end{table}

 This section first verifies the developed analytical model via independent system-level Monte Carlo simulations. Then, numerical results are presented to highlight the tradeoff between data granularity, reliability, and delay. The numerical results also demonstrate the effect of power control and antenna directivity. {For the sake of fair comparison between the constant power transmission model and the power control model, the total power consumption is set to be equal in both models\footnote{ {The values of $P$ and $\rho$ utilized are within the practical ranges of IoT systems such as LoRA and Sigfox \cite{sundaram2019survey}. The low transmission power and high receiver sensitivity is essential for IoT devices with strict power budget.}}, such that $P = \frac{\rho}{ \mathcal{N}_G} \sum_{i=1}^Y \sum_{j} \mathbbm{1}_{\{ v_{ij} \in \Psi_i \cap \; \tinyvarhexagon_\circ\}} \left\|v_{ij} \right\|^\eta$.
 Unless otherwise stated, the results in this section implement the network parameters given in Table 1.}

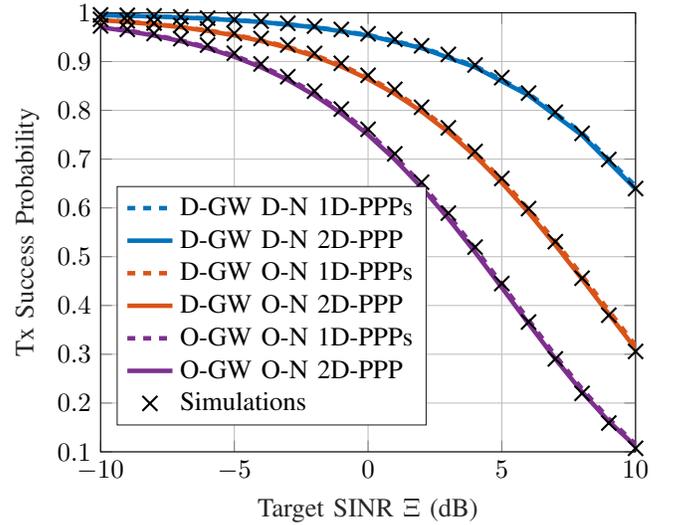
\begin{figure}[!t]
\centering
\begin{tikzpicture}
%
%
\definecolor{mycolor1}{rgb}{0.00000,0.44700,0.74100}%
\definecolor{mycolor2}{rgb}{0.85000,0.32500,0.09800}%
\definecolor{mycolor3}{rgb}{0.92900,0.69400,0.12500}%
\definecolor{mycolor4}{rgb}{0.49400,0.18400,0.55600}%
\definecolor{mycolor5}{rgb}{0.46600,0.67400,0.18800}%
\definecolor{mycolor6}{rgb}{0.30100,0.74500,0.93300}%
\definecolor{mycolor7}{rgb}{0.63500,0.07800,0.18400}%

\begin{axis}[%
width=2.8in,
height=2.3in,
at={(0.1158in,0.081in)},
scale only axis,
xmin=-10,
xmax=10,
xlabel style={font=\color{white!15!black}},
xlabel={Target SINR $\Xi$ (dB)},
ymin=0.1,
ymax=1,
ytick={0,0.1,...,1},
xmajorgrids,
ymajorgrids,
ylabel style={font=\color{white!15!black}},
ylabel={Tx Success Probability},
axis background/.style={fill=white},
legend style={font=\footnotesize,at={(0.03,0.06)}, anchor=south west, legend cell align=left, align=left, draw=white!15!black}
]

\addplot [dashed, line width=0.6mm, color=mycolor1]
  table[row sep=crcr]{%
-10	0.9953\\
-8	0.9926\\
-6	0.9883\\
-4	0.9815\\
-2	0.9709\\
0	0.9545\\
2	0.9291\\
4	0.8907\\
6	0.8341\\
8	0.7534\\
10	0.6445\\
};
\addlegendentry{\normalsize{D-GW D-N 1D-PPPs}}

\addplot [line width=0.6mm, color=mycolor1]
  table[row sep=crcr]{%
-10	0.9952\\
-8	0.9924\\
-6	0.988\\
-4	0.9811\\
-2	0.9703\\
0	0.9534\\
2	0.9276\\
4	0.8884\\
6	0.8308\\
8	0.7489\\
10	0.6388\\
};
\addlegendentry{\normalsize{D-GW D-N 2D-PPP}}

\addplot [dashed, line width=0.6mm, color=mycolor2]
  table[row sep=crcr]{%
-10	0.9853526094363\\
-9	0.981615845187336\\
-8	0.976943530097882\\
-7	0.97111150525732\\
-6	0.963847542412016\\
-5	0.954824167817079\\
-4	0.943652205068258\\
-3	0.929876299607367\\
-2	0.9129743080193\\
-1	0.892363163450566\\
0	0.867414554789608\\
1	0.837484259022529\\
2	0.80195888411659\\
3	0.760322641715873\\
4	0.712244101097598\\
5	0.6576784209491\\
6	0.59697455396291\\
7	0.53097032220558\\
8	0.461052704403035\\
9	0.389158141874109\\
10	0.317690001367161\\
};
\addlegendentry{\normalsize{D-GW O-N 1D-PPPs}}

\addplot [line width=0.6mm, color=mycolor2]
  table[row sep=crcr]{%
-10	0.984940089104847\\
-9	0.981100670514556\\
-8	0.976301459506261\\
-7	0.970313299057502\\
-6	0.962858333898829\\
-5	0.953602986639979\\
-4	0.942151814167623\\
-3	0.928043572291715\\
-2	0.910751437356818\\
-1	0.889690040455098\\
0	0.864232638684383\\
1	0.833742143842552\\
2	0.797619495237283\\
3	0.755371562035889\\
4	0.706697995901487\\
5	0.651592075460276\\
6	0.590444901976027\\
7	0.524136209508668\\
8	0.454089929123004\\
9	0.382270147201527\\
10	0.311094941719484\\
};
\addlegendentry{\normalsize{D-GW O-N 2D-PPP}}

\addplot [dashed, line width=0.6mm, color=mycolor4]
  table[row sep=crcr]{%
-10	0.970961903701639\\
-9	0.963635601207222\\
-8	0.954521490445355\\
-7	0.94321727656902\\
-6	0.929248920563159\\
-5	0.912068265307471\\
-4	0.891057129686988\\
-3	0.865541766610479\\
-2	0.834822540936486\\
-1	0.798224148291185\\
0	0.755170979024282\\
1	0.705289418130788\\
2	0.648533067314879\\
3	0.585317760664419\\
4	0.516641889083368\\
5	0.444157108629128\\
6	0.370150123365897\\
7	0.297403447401331\\
8	0.228924911720534\\
9	0.167569966503077\\
10	0.115619379964675\\
};
\addlegendentry{\normalsize{O-GW O-N 1D-PPPs}}

\addplot [line width=0.6mm, color=mycolor4]
  table[row sep=crcr]{%
-10	0.970153494120293\\
-9	0.962631259428523\\
-8	0.953277882656906\\
-7	0.941683771884944\\
-6	0.927367629805935\\
-5	0.90977490155846\\
-4	0.888283127367683\\
-3	0.862218168622368\\
-2	0.830886147621605\\
-1	0.793626229249444\\
0	0.749888378390371\\
1	0.699337074133658\\
2	0.641975888543252\\
3	0.578278762722107\\
4	0.509303064485051\\
5	0.436750310296672\\
6	0.362937516084961\\
7	0.290650165996855\\
8	0.222868976660291\\
9	0.162394904612737\\
10	0.11143348292807\\
};
\addlegendentry{\normalsize{O-GW O-N 2D-PPP}}

\addplot [color=white, draw=none, mark=x, thick, mark size=4pt, mark options={ black}]
  table[row sep=crcr]{%
-10	0.9957\\
-9	0.99484\\
-8	0.99356\\
-7	0.99152\\
-6	0.98898\\
-5	0.98608\\
-4	0.98282\\
-3	0.97848\\
-2	0.97264\\
-1	0.96578\\
0	0.95698\\
1	0.94562\\
2	0.93232\\
3	0.91454\\
4	0.89292\\
5	0.86704\\
6	0.83534\\
7	0.79626\\
8	0.75276\\
9	0.69928\\
10	0.63966\\
};
\addlegendentry{\normalsize{Simulations}}

\addplot [color=black, draw=none, mark=x, thick, mark size=4pt, mark options={ black}]
  table[row sep=crcr]{%
-10	0.9867\\
-9	0.98332\\
-8	0.97894\\
-7	0.97406\\
-6	0.96712\\
-5	0.95766\\
-4	0.94738\\
-3	0.93404\\
-2	0.91674\\
-1	0.89638\\
0	0.87178\\
1	0.84304\\
2	0.80626\\
3	0.7641\\
4	0.7157\\
5	0.6604\\
6	0.59864\\
7	0.53086\\
8	0.45594\\
9	0.38004\\
10	0.3058\\
};

\addplot [color=black, draw=none, mark=x, thick, mark size=4pt, mark options={ black}]
  table[row sep=crcr]{%
-10	0.97306\\
-9	0.96598\\
-8	0.95746\\
-7	0.9469\\
-6	0.93382\\
-5	0.91704\\
-4	0.8951\\
-3	0.86896\\
-2	0.83938\\
-1	0.80258\\
0	0.7612\\
1	0.71084\\
2	0.65262\\
3	0.58932\\
4	0.51968\\
5	0.44496\\
6	0.36612\\
7	0.29054\\
8	0.21984\\
9	0.159\\
10	0.1072\\
};

\end{axis}
\end{tikzpicture}
    \caption{{Success probability at $r_o$ = 300 m for all antenna directivity scenarios.}}
    \label{ant}
\end{figure}

Fig. \ref{ant} shows the transmission success probability $\mathbb{P}\{\text{SINR}>\Xi\}$ for the 1D-PPP approximation, the 2D-PPP approximation, and the exact Monte Carlo simulations for the constant Tx power scheme with different antenna directivity scenarios. The close match between the analysis and simulations verifies both the 1D-PPP and the 2D-PPP approximations for all antenna directivity scenarios. Note that the D-GW with D-N scenario in Fig. \ref{ant} further verifies Approximation 1, which assumes a random orientation for directive devices. Since both approximations are accurate, the 2D-PPP approximation is preferred due to its analytical simplicity when compared to the 1D-PPP approximation.  

In addition to verifying the mathematical model, Fig. \ref{ant} also demonstrates the significant performance improvement offered by the directive antenna. For instance, for a $90\%$ target Tx success probability, equipping the gateways with directive antennas offers $3$ dB SINR gain when compared to the omni-gateways. Such gain could be increased to {$8$} dB if the directive antennas are implemented in both the gateways and devices. Note that the improved target SINR $\Xi$ for the same transmission reliability translates to higher transmission rates and less packet delay.

Fig. \ref{sc_pr} confirms the accuracy of the 2D-PPP approximation as well as the improved performance of the directive antenna in the power control scheme. Furthermore, the D-GW with D-N scenario verifies Approximation 1 for the power control scheme. The SINR improvements harvested from the directive antenna in the power control scheme are approximately equal to that of the constant power scheme shown in Fig.~\ref{ant}.

\begin{figure*}[ht!]
   \subfloat[\label{sys_geo}]{%
%
%
\definecolor{mycolor1}{rgb}{0.00000,0.44700,0.74100}%
\definecolor{mycolor2}{rgb}{0.85000,0.32500,0.09800}%
\begin{tikzpicture}

\begin{axis}[%
width=2in,
height=2in,
at={(0.758in,0.481in)},
scale only axis,
xmin=-10,
xmax=10,
xlabel style={font=\color{white!15!black}},
xlabel={Target SINR $\Xi$ (dB)},
ymin=0,
ymax=1,
xmajorgrids,
ymajorgrids,
ylabel style={font=\color{white!15!black}},
ylabel={Tx Success probability},
axis background/.style={fill=white},
legend style={at={(0.05,0.1)}, anchor=south west, legend cell align=left, align=left, draw=white!15!black}
]
\addplot [line width=0.6mm, color=mycolor1]
 plot [error bars/.cd, y dir = both, y explicit]
 table[row sep=crcr, y error plus index=2, y error minus index=3]{%
-10	0.993364358719591	0.0061312200366277	0.0061312200366277\\
-8	0.98953541090247	0.0096466376050068	0.0096466376050068\\
-6	0.9835436593191	0.0151150431944723	0.0151150431944723\\
-4	0.974233964299924	0.023533084163614	0.023533084163614\\
-2	0.959925271095767	0.0362873280944676	0.0362873280944676\\
0	0.938288560565576	0.0551563778814547	0.0551563778814547\\
2	0.906343440485202	0.0821120762481949	0.0821120762481949\\
4	0.860762679839403	0.118733381624225	0.118733381624225\\
6	0.798737951897565	0.165098787186187	0.165098787186187\\
8	0.719541680068747	0.218386770201768	0.218386770201768\\
10	0.626379123880381	0.272182941702383	0.272182941702383\\
};
\addlegendentry{D-GW D-N}

\addplot[dashed, line width=0.6mm, color=mycolor2]
  table[row sep=crcr]{%
-10	0.992219810445793\\
-9	0.990226111934526\\
-8	0.987728078363195\\
-7	0.984601812479596\\
-6	0.980695003201365\\
-5	0.975821491322651\\
-4	0.969755362141904\\
-3	0.962224836731949\\
-2	0.952906426870754\\
-1	0.941420072041406\\
0	0.927326289316434\\
1	0.910126729061787\\
2	0.889269923875721\\
3	0.864164420598085\\
4	0.834201860747352\\
5	0.79879286094519\\
6	0.757418610977233\\
7	0.709700687868663\\
8	0.655490205440643\\
9	0.594974358455221\\
10	0.528792789341669\\
};
\addlegendentry{D-GW D-N PC}

\end{axis}
\end{tikzpicture}%
}
   \subfloat[\label{intr_field} ]{%
%
%
\definecolor{mycolor1}{rgb}{0.00000,0.44700,0.74100}%
\definecolor{mycolor2}{rgb}{0.85000,0.32500,0.09800}%
\begin{tikzpicture}

\begin{axis}[%
width=2in,
height=2in,
at={(0.758in,0.481in)},
scale only axis,
xmin=-10,
xmax=10,
xlabel style={font=\color{white!15!black}},
xlabel={Target SINR $\Xi$ (dB)},
ymin=0,
ymax=1,
xmajorgrids,
ymajorgrids,
ylabel style={font=\color{white!15!black}},
ylabel={Tx Success probability},
axis background/.style={fill=white},
legend style={at={(0.7,0.3)},legend cell align=left, align=left, draw=white!15!black}
]
\addplot [line width=0.6mm, color=mycolor1]
 plot [error bars/.cd, y dir = both, y explicit]
 table[row sep=crcr, y error plus index=2, y error minus index=3]{%
-10	0.979448683177472	0.0187893789396509	0.0187893789396509\\
-8	0.967997086267789	0.0290240677258021	0.0290240677258021\\
-6	0.950626544355397	0.0442367887451087	0.0442367887451087\\
-4	0.924852909794198	0.0661353771049909	0.0661353771049909\\
-2	0.887779503218107	0.0962733587616261	0.0962733587616261\\
0	0.836639849389617	0.135312236607978	0.135312236607978\\
2	0.769815258250878	0.182050136553879	0.182050136553879\\
4	0.688163081424879	0.23269875162496	0.23269875162496\\
6	0.596055509796716	0.281144154760236	0.281144154760236\\
8	0.501186500685785	0.320619262998541	0.320619262998541\\
10	0.412477945303459	0.346111242347246	0.346111242347246\\
};
\addlegendentry{D-GW O-N}

\addplot [dashed, line width=0.6mm, color=mycolor2]
  table[row sep=crcr]{%
-10	0.97137872458115\\
-9	0.964276156119545\\
-8	0.955501963406373\\
-7	0.944709805937927\\
-6	0.931504638064954\\
-5	0.915446637320759\\
-4	0.896060815648494\\
-3	0.872853850053224\\
-2	0.845339295172258\\
-1	0.813071610168658\\
0	0.775688414874376\\
1	0.732959258948577\\
2	0.684838145458093\\
3	0.631516196351029\\
4	0.573470072691151\\
5	0.511500766343623\\
6	0.446755847523069\\
7	0.380726144991245\\
8	0.315205690438331\\
9	0.25220291249308\\
10	0.193793692311859\\
};
\addlegendentry{D-GW O-N PC}

\end{axis}
\end{tikzpicture}%
}
   \subfloat[\label{intr_field_arae}]{%
%
%
\definecolor{mycolor1}{rgb}{0.00000,0.44700,0.74100}%
\definecolor{mycolor2}{rgb}{0.85000,0.32500,0.09800}%
\begin{tikzpicture}

\begin{axis}[%
width=2in,
height=2in,
at={(0.758in,0.481in)},
scale only axis,
xmin=-10,
xmax=10,
xlabel style={font=\color{white!15!black}},
xlabel={Target SINR $\Xi$ (dB)},
ymin=0,
ymax=1,
xmajorgrids,
ymajorgrids,
ylabel style={font=\color{white!15!black}},
ylabel={Tx Success probability},
axis background/.style={fill=white},
legend style={at={(0.7,0.3)},legend cell align=left, align=left, draw=white!15!black}
]
\addplot [line width=0.6mm, color=mycolor1]
 plot [error bars/.cd, y dir = both, y explicit]
 table[row sep=crcr, y error plus index=2, y error minus index=3]{%
-10	0.959822937958891	0.036346463048475	0.036346463048475\\
-8	0.938205674920674	0.055143360338662	0.055143360338662\\
-6	0.906404330280799	0.0818426919959294	0.0818426919959294\\
-4	0.861281512334024	0.117800117211014	0.117800117211014\\
-2	0.800344177176789	0.162809336608582	0.162809336608582\\
0	0.723177219298036	0.21397440919669	0.21397440919669\\
2	0.632911094678294	0.265450684714918	0.265450684714918\\
4	0.536634249341669	0.309921003961972	0.309921003961972\\
6	0.443737938262148	0.341334847512383	0.341334847512383\\
8	0.362444743065868	0.356854886621963	0.356854886621963\\
10	0.296551367530121	0.356591436513425	0.356591436513425\\
};
\addlegendentry{O-Gw O-N}

\addplot [dashed, line width=0.6mm, color=mycolor2]
  table[row sep=crcr]{%
-10	0.943931874179444\\
-9	0.930371679767341\\
-8	0.913807487868227\\
-7	0.893712282198062\\
-6	0.869532236200165\\
-5	0.840717021477718\\
-4	0.806763479916699\\
-3	0.767272500137497\\
-2	0.722016290549924\\
-1	0.671010119946092\\
0	0.614579776555315\\
1	0.553414330825562\\
2	0.488593742009962\\
3	0.421582182850649\\
4	0.354179911711506\\
5	0.28842855604579\\
6	0.226467288181167\\
7	0.170342375882429\\
8	0.121782317745266\\
9	0.0819663145862249\\
10	0.0513325745199281\\
};
\addlegendentry{O-Gw O-N PC}

\end{axis}
\end{tikzpicture}%
}
\caption{{\label{eff_pc}Effect of power control scheme on the success probability with error bars representing the standard deviation for the constant transmission power relative to its mean value (blue line).}}
\end{figure*}
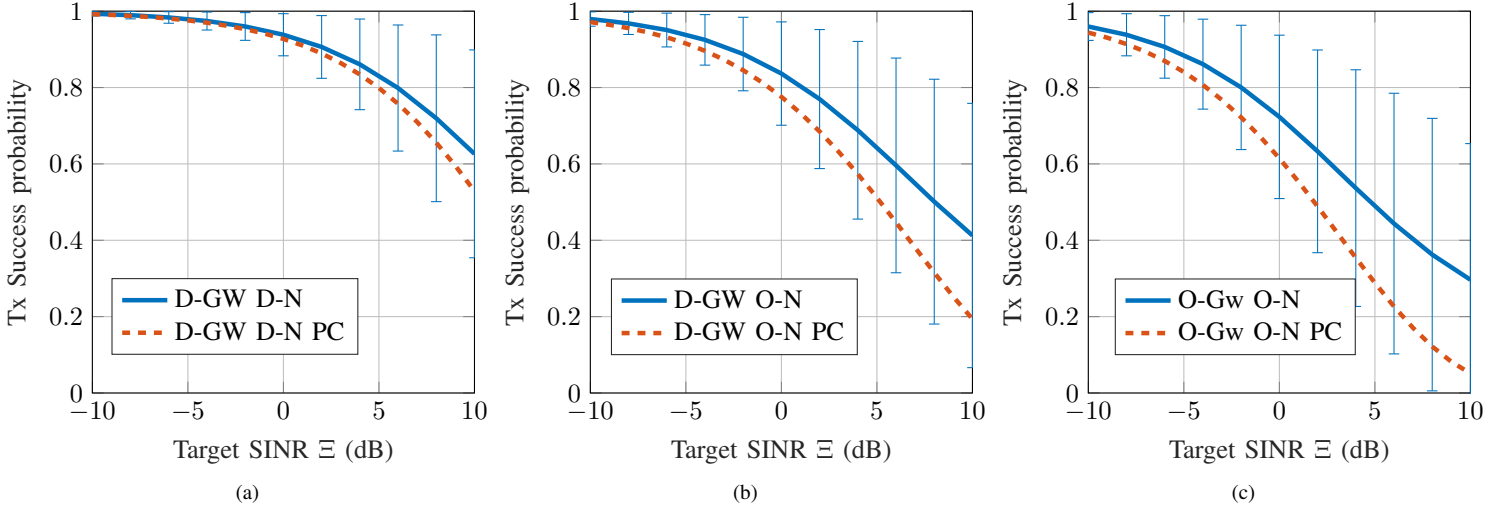

 {Fig.~\ref{eff_pc} shows the success probability  of channel inversion power control when compared to the mean  success probability of the constant Tx power scheme with  error bars highlighting the high standard deviation among the cell-center and cell-edge devices.
In general and for the same total power consumption, there is a trade off between the average performance and fairness. The Tx success probability and consequently throughput for  power control is always lower than the average Tx success  probability and average throughput in case  of  constant  Tx  power. On the other hand, the channel inversion power control improves the fairness between devices by providing a  unified  Tx success  probability and throughput for all devices regardless of their location. Note that the gap between the average performance of the constant Tx power and the power control scheme decreases for directive antennas. It is worth mentioning that in power control model the battery lifetime for devices at the edge of the cell is lower than devices near to cell centre. However, the battery lifetime is the same for all devices in the case of constant power transmission.}

\begin{figure}[!]
\centering
%
%
\definecolor{mycolor1}{rgb}{0.00000,0.44700,0.74100}%
\definecolor{mycolor2}{rgb}{0.85000,0.32500,0.09800}%
\definecolor{mycolor3}{rgb}{0.92900,0.69400,0.12500}%
\definecolor{mycolor4}{rgb}{0.49400,0.18400,0.55600}%
\definecolor{mycolor5}{rgb}{0.46600,0.67400,0.18800}%
\definecolor{mycolor6}{rgb}{0.30100,0.74500,0.93300}%
\definecolor{mycolor7}{rgb}{0.63500,0.07800,0.18400}%
\begin{tikzpicture}

\begin{axis}[%
width=2.8in,
height=2.3in,
at={(0.758in,0.481in)},
scale only axis,
xmin=-10,
xmax=10,
xlabel style={font=\color{white!15!black}},
xlabel={Target SINR $\Xi$ (dB)},
ymin=0,
ymax=1,
ytick={0,0.1,...,1},
xmajorgrids,
ymajorgrids,
ylabel style={font=\color{white!15!black}},
ylabel={Tx Success Probability},
axis background/.style={fill=white},
legend style={at={(0.03,0.06)}, anchor=south west, legend cell align=left, align=left, draw=white!15!black}
]

\addplot [line width=0.6mm, color=mycolor1]
  table[row sep=crcr]{%
-10	0.992219810445793\\
-9	0.990226111934525\\
-8	0.987728078363195\\
-7	0.984601812479596\\
-6	0.980695003201365\\
-5	0.975821491322651\\
-4	0.969755362141904\\
-3	0.962224836731949\\
-2	0.952906426870754\\
-1	0.941420072041406\\
0	0.927326289316434\\
1	0.910126729061787\\
2	0.889269923875721\\
3	0.864164420598085\\
4	0.834201860747352\\
5	0.79879286094519\\
6	0.757418610977233\\
7	0.709700687868663\\
8	0.655490205440643\\
9	0.594974358455221\\
10	0.528792789341669\\
};
\addlegendentry{D-GW D-N}

\addplot [line width=0.6 mm, color=mycolor2]
  table[row sep=crcr]{%
-10	0.97137872458115\\
-9	0.964276156119545\\
-8	0.955501963406373\\
-7	0.944709805937927\\
-6	0.931504638064954\\
-5	0.915446637320759\\
-4	0.896060815648494\\
-3	0.872853850053224\\
-2	0.845339295172258\\
-1	0.813071610168657\\
0	0.775688414874376\\
1	0.732959258948577\\
2	0.684838145458093\\
3	0.631516196351029\\
4	0.573470072691151\\
5	0.511500766343623\\
6	0.446755847523069\\
7	0.380726144991245\\
8	0.315205690438331\\
9	0.25220291249308\\
10	0.193793692311859\\
};
\addlegendentry{D-GW O-N}

\addplot [line width=0.6mm, color=mycolor4]
  table[row sep=crcr]{%
-10	0.943931874179444\\
-9	0.930371679767341\\
-8	0.913807487868227\\
-7	0.893712282198062\\
-6	0.869532236200165\\
-5	0.840717021477718\\
-4	0.806763479916699\\
-3	0.767272500137497\\
-2	0.722016290549924\\
-1	0.671010119946092\\
0	0.614579776555315\\
1	0.553414330825562\\
2	0.488593742009962\\
3	0.421582182850649\\
4	0.354179911711506\\
5	0.28842855604579\\
6	0.226467288181167\\
7	0.170342375882429\\
8	0.121782317745266\\
9	0.0819663145862249\\
10	0.0513325745199281\\
};
\addlegendentry{O-GW O-N}

\addplot [color=white, draw=none, mark=x,  thick, mark size=4 pt, mark options={ black}]
  table[row sep=crcr]{%
-10	0.99224\\
-9	0.9902\\
-8	0.98784\\
-7	0.98462\\
-6	0.98128\\
-5	0.97618\\
-4	0.97048\\
-3	0.96308\\
-2	0.95362\\
-1	0.94312\\
0	0.92914\\
1	0.9122\\
2	0.89212\\
3	0.86692\\
4	0.83576\\
5	0.79928\\
6	0.75634\\
7	0.70584\\
8	0.6492\\
9	0.58478\\
10	0.5151\\
};
\addlegendentry{Simulations}

\addplot [color=mycolor2, draw=none, mark=x, thick, mark size=4 pt, mark options={ black}]
  table[row sep=crcr]{%
-10	0.97368\\
-9	0.96708\\
-8	0.9585\\
-7	0.94878\\
-6	0.9371\\
-5	0.92174\\
-4	0.90346\\
-3	0.88186\\
-2	0.85614\\
-1	0.82526\\
0	0.7894\\
1	0.74854\\
2	0.70122\\
3	0.64798\\
4	0.592\\
5	0.52956\\
6	0.4618\\
7	0.3951\\
8	0.32826\\
9	0.26118\\
10	0.1986\\
};
\addplot [color=mycolor2, draw=none, mark=x, thick, mark size=4 pt, mark options={ black}]
  table[row sep=crcr]{%
-10	0.94856\\
-9	0.93612\\
-8	0.92064\\
-7	0.90316\\
-6	0.88054\\
-5	0.85326\\
-4	0.8201\\
-3	0.78234\\
-2	0.7386\\
-1	0.6898\\
0	0.63542\\
1	0.5744\\
2	0.50884\\
3	0.44014\\
4	0.37004\\
5	0.30156\\
6	0.23712\\
7	0.1775\\
8	0.12586\\
9	0.08588\\
10	0.0528\\
};

\end{axis}
\end{tikzpicture}%
    \caption{{Success probability with path-loss inversion power control for all antenna directivity scenarios.}}
    \label{sc_pr}
\end{figure}
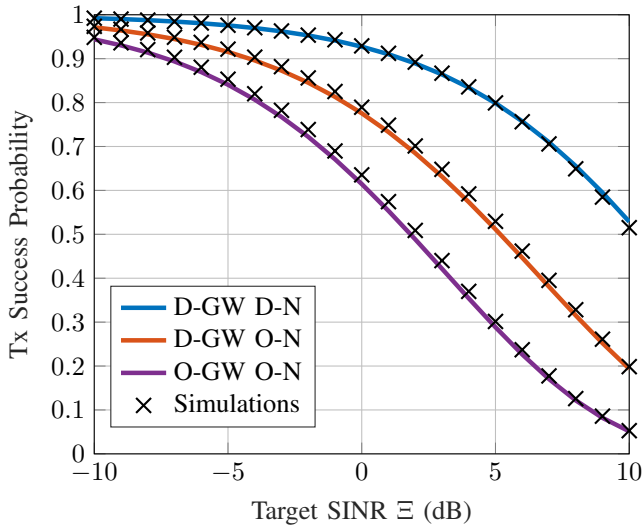

\begin{figure}[!t]
\centering

%
%
\definecolor{mycolor1}{rgb}{0.00000,0.44700,0.74100}%
\definecolor{mycolor2}{rgb}{0.85000,0.32500,0.09800}%
\definecolor{mycolor3}{rgb}{0.92900,0.69400,0.12500}%
\definecolor{mycolor4}{rgb}{0.49400,0.18400,0.55600}%

\begin{tikzpicture}

\begin{axis}[%
width=2.8in,
height=2.3in,
at={(0.758in,0.481in)},
scale only axis,
xmin=-10,
xmax=20,
xlabel style={font=\color{white!15!black}},
xlabel={Target SINR $\Xi$ (dB)},
ymin=0,
ymax=1.8,
ytick={0,0.2,...,1.8},
xmajorgrids,
ymajorgrids,
ylabel style={font=\color{white!15!black}},
ylabel={Throughput $\mathcal{T} $(nats/sec/Hz)},
axis background/.style={fill=white},
legend style={at={(0.015,0.75)}, anchor=south west, legend cell align=left, align=left, draw=white!15!black}
]
\addplot [line width=0.6mm, color=mycolor1]
  table[row sep=crcr]{%
-10	0.0945686485390017\\
-9	0.117417141600987\\
-8	0.14531143898166\\
-7	0.179125330744603\\
-6	0.219767879779272\\
-5	0.268126366180476\\
-4	0.324983738891472\\
-3	0.390909886668997\\
-2	0.46613054040363\\
-1	0.550383084561538\\
0	0.642773602898802\\
1	0.741652459163157\\
2	0.844525586377893\\
3	0.948015860219639\\
4	1.04788545192735\\
5	1.13912849545969\\
6	1.2161454584893\\
7	1.27301584114528\\
8	1.30389123172265\\
9	1.3035309133393\\
10	1.26798972985225\\
11	1.19543422185145\\
12	1.08700090996164\\
13	0.947520163160814\\
14	0.785828556287283\\
15	0.614331046228241\\
16	0.447535021622426\\
17	0.299550749095873\\
18	0.181051200914246\\
19	0.0967322043295168\\
20	0.0445054330951122\\
};
\addlegendentry{D-GW D-N}

\addplot [line width=0.6mm, color=mycolor2]
  table[row sep=crcr]{%
-10	0.0925822808979253\\
-9	0.114340097277732\\
-8	0.140570434610371\\
-7	0.171867910765001\\
-6	0.20874461340562\\
-5	0.251537174042206\\
-4	0.300287273999087\\
-3	0.354602361711803\\
-2	0.413512230972195\\
-1	0.475346632246396\\
0	0.537666237763187\\
1	0.597280597863546\\
2	0.650379958703937\\
3	0.69279335720858\\
4	0.720366346042925\\
5	0.729432030353952\\
6	0.717331323979112\\
7	0.682922282585871\\
8	0.627002406047148\\
9	0.552552035557478\\
10	0.464696978692748\\
11	0.37029756589851\\
12	0.277118828686465\\
13	0.192639129091309\\
14	0.122708510078328\\
15	0.0704257951757443\\
16	0.0356647030147735\\
17	0.0155289663627037\\
18	0.0056297098329093\\
19	0.00163284074341889\\
20	0.00036057259632374\\
};
\addlegendentry{D-GW O-N}

\addplot [line width=0.6mm, color=mycolor4]
  table[row sep=crcr]{%
-10	0.0899663166510762\\
-9	0.110319837003059\\
-8	0.134436474899442\\
-7	0.162590101003455\\
-6	0.194856969114384\\
-5	0.231003725537297\\
-4	0.270362013286871\\
-3	0.311709274821536\\
-2	0.353186665766825\\
-1	0.39229312243907\\
0	0.425994239348478\\
1	0.450971371663832\\
2	0.464009751587189\\
3	0.462489065971755\\
4	0.444904278342057\\
5	0.411317130084612\\
6	0.36362608473875\\
7	0.30554929229093\\
8	0.242247549954112\\
9	0.179580217865271\\
10	0.123090137781906\\
11	0.0769386329371214\\
12	0.0431100597615686\\
13	0.0211983621019391\\
14	0.00891011371700252\\
15	0.00309845997977363\\
16	0.000856058721992455\\
17	0.00017869916126771\\
18	2.64773527198414e-05\\
19	2.57646141065435e-06\\
20	1.49478311680235e-07\\
};
\addlegendentry{O-GW O-N}

\end{axis}
\end{tikzpicture}%
    \caption{{Throughput ($\mathcal{T}$) with power control. }}
    \label{Outage_capacity}
\end{figure}
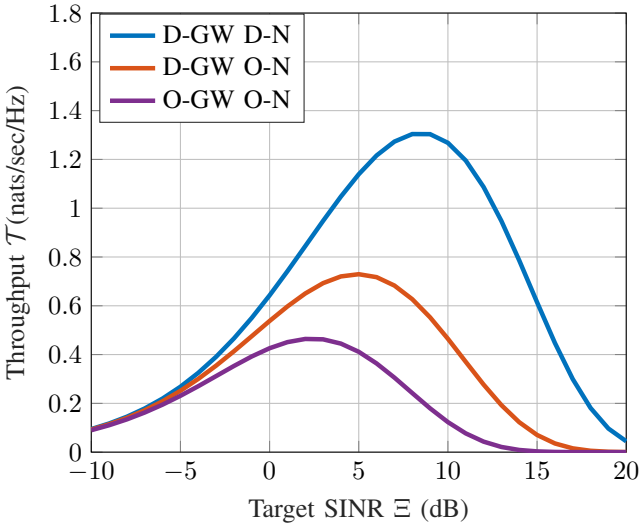

Fig.~\ref{Outage_capacity} plots the throughput versus the target SINR $\Xi$ for the power control scheme, which demonstrates the Tx throughput improvements offered by directive antennas. The figure shows a delicate tradeoff between channel utilization and transmission reliability. Operating at low $\Xi$ leads to high Tx reliability but low uplink channel utilization. Increasing $\Xi$ improves channel utilization at the expense of decreasing Tx reliability. Hence, there is an optimal target SINR $\Xi^*$ that balances the tradeoff between reliability and channel utilization. When compared to the primitive O-GW with O-D, the figure quantifies the throughput improvement offered by directive antennas as {$55\%$} and {$180\%$} for D-GW with O-N and D-GW with D-N, respectively.

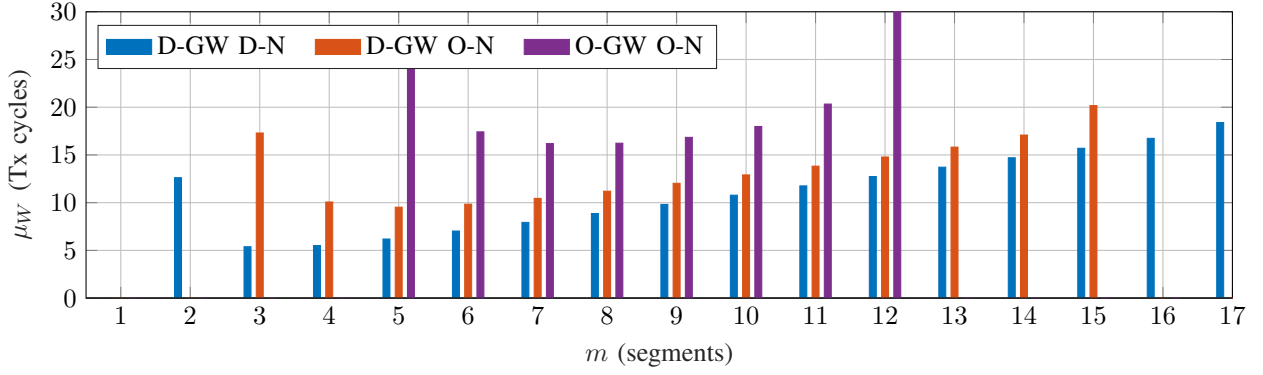
\begin{figure*}[ht!]
\centering
%
%
\definecolor{mycolor1}{rgb}{0.00000,0.44700,0.74100}%
\definecolor{mycolor2}{rgb}{0.85000,0.32500,0.09800}%
\definecolor{mycolor3}{rgb}{0.92900,0.69400,0.12500}%
\definecolor{mycolor4}{rgb}{0.49400,0.18400,0.55600}%

\begin{tikzpicture}

\begin{axis}[%
width=6in,
height=1.5in,
at={(0.758in,0.481in)},
scale only axis,
bar shift auto,
xmin=0.5,
xmax=17,
xlabel style={font=\color{white!15!black}},
xlabel={$m$ (segments)},
ymin=0,
ymax=30,
ytick={0,5,...,30},
xmajorgrids,
ymajorgrids,
ylabel style={font=\color{white!15!black}},
ylabel={$\mu_W$ (Tx cycles)},
xtick={1,...,17},
axis background/.style={fill=white},
legend style={at={(0.01,0.8)}, anchor=south west, legend columns=3, legend cell align=left, align=left, draw=white!15!black}
]
\addplot[ybar, bar width=0.1, fill=mycolor1, draw=mycolor1, area legend] table[row sep=crcr] {%
1	0\\
2	12.6166972755635\\
3	5.38449089321553\\
4	5.51245235084528\\
5	6.19269300592851\\
6	7.0294348937334\\
7	7.93101735341328\\
8	8.86542516731135\\
9	9.81866857076195\\
10	10.7836925620928\\
11	11.7565672118953\\
12	12.7349539278891\\
13	13.7174947217065\\
14	14.7040504297133\\
15	15.6990279401275\\
16	16.7359685653286\\
17	18.3919892753271\\
18	0\\
};
\addplot[forget plot, color=white!15!black] table[row sep=crcr] {%
0.1	0\\
17	0\\
};
\addlegendentry{D-GW D-N $\;\;$}

\addplot[ybar, bar width=0.1, fill=mycolor2, draw=mycolor2, area legend] table[row sep=crcr] {%
1	0\\
2	0\\
3	17.2989372146269\\
4	10.0713526576663\\
5	9.528080064795\\
6	9.84030451859126\\
7	10.4508749807064\\
8	11.2015203659221\\
9	12.0300639985623\\
10	12.9084811441899\\
11	13.8255811813508\\
12	14.7849321728999\\
13	15.8198677177728\\
14	17.0855240600075\\
15	20.173828921529\\
16	0\\
17	0\\
18	0\\
};
\addplot[forget plot, color=white!15!black] table[row sep=crcr] {%
0.1	0\\
17	0\\
};
\addlegendentry{D-GW O-N $\;\;$}

\addplot[ybar, bar width=0.1, fill=mycolor4, draw=mycolor4, area legend] table[row sep=crcr] {%
1	0\\
2	0\\
3	0\\
4	0\\
5	26.5142144970981\\
6	17.417478013887\\
7	16.1879134388073\\
8	16.2294092530317\\
9	16.8419872409174\\
10	17.9828835508534\\
11	20.3288384450355\\
12	36.371853507558\\
13	0\\
14	0\\
15	0\\
16	0\\
17	0\\
18	0\\
};
\addplot[forget plot, color=white!15!black] table[row sep=crcr] {%
0.1	0\\
17	0\\
};
\addlegendentry{O-GW O-N}

\end{axis}
\end{tikzpicture}%
    \caption{ {Average packet delay for the power control scheme at $L=80$ kbits. Infinite delays cannot be presented and hence omitted from the figure.}}
     \label{td_1_100}
\end{figure*}

The results in Fig.~\ref{Outage_capacity} motivate large packet segmentation to avoid uplink over-utilization and the subsequent low Tx reliability. In this regard, Fig.~\ref{td_1_100} plots the average packet delay $(\mu_W)$ to demonstrate the necessity of packet segmentation. Note that the average delay is given in the number of Tx cycles, where each Tx cycle is $T_c =\mathcal{N}_G \times T_s$ seconds. Since infinite delay cannot be presented, packet segmentation $m$ that leads to unstable queueing system (i.e., $\rho_m \geq 1$) are omitted from the figure. The figure shows that there is an optimal packet segmentation $m^*$ that minimizes the delay for each of the antenna schemes. For instance, the D-GW with D-N delay is minimized at $m^*=3$ segments per packet. The required number of segments that minimizes the delay increases to $m^*=5$ for the D-GW with O-N and {$m^*=7$} for the O-GW with O-N. The D-GW with D-N offers the least delay because of its higher throughput (see Fig.~\ref{Outage_capacity}). As the throughput decreases, the packet has to be divided into a large number of smaller segments to avoid the low-reliability regime. Maintaining the Tx reliability by increasing the number of segments per packet leads to a larger delay. 


 
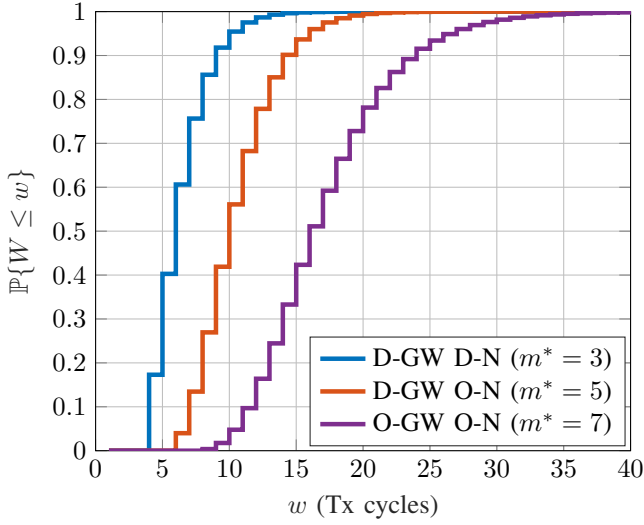
\begin{figure}[!h]
\centering
%
%
\definecolor{mycolor1}{rgb}{0.00000,0.44700,0.74100}%
\definecolor{mycolor2}{rgb}{0.85000,0.32500,0.09800}%
\definecolor{mycolor3}{rgb}{0.92900,0.69400,0.12500}%
\definecolor{mycolor4}{rgb}{0.49400,0.18400,0.55600}%
\begin{tikzpicture}

\begin{axis}[%
width=2.8in,
height=2.3in,
at={(0.758in,0.481in)},
scale only axis,
xmin=0,
xmax=40,
xlabel style={font=\color{white!15!black}},
xlabel={$w$ (Tx cycles)},
ymin=0,
ymax=1,
xtick={0,5,...,40},
ytick={0,0.1,...,1},
ylabel style={font=\color{white!15!black}},
ylabel={$\mathbb{P}\{W\leq w\}$},
xmajorgrids,
ymajorgrids,
axis background/.style={fill=white},
legend style={at={(0.4,0.01)}, anchor=south west, legend cell align=left, align=left, legend columns=1, draw=white!15!black}
]
\addplot[const plot,line width=0.6mm, color=mycolor1] table[row sep=crcr] {%
1	0\\
2	0\\
3	0\\
4	0.172955516391443\\
5	0.402720576038063\\
6	0.606210212124346\\
7	0.756392721953493\\
8	0.85614869981625\\
9	0.917992373370275\\
10	0.954506618675465\\
11	0.975295691063742\\
12	0.986802991122137\\
13	0.99303103877148\\
14	0.996340531916284\\
15	0.998072502319032\\
16	0.998967280878403\\
17	0.999424465684677\\
18	0.999655838349796\\
19	0.999771955942764\\
20	0.999829802678946\\
21	0.999858432053017\\
22	0.99987251837648\\
23	0.999879412700075\\
24	0.999882770947281\\
25	0.99988439967886\\
26	0.999885186482554\\
27	0.999885565192645\\
28	0.999885746868535\\
29	0.999885833754447\\
30	0.999885875188971\\
31	0.999885894896171\\
32	0.999885904246275\\
33	0.999885908672245\\
34	0.999885910762818\\
35	0.999885911748294\\
36	0.999885912211958\\
37	0.999885912429722\\
38	0.999885912531825\\
39	0.999885912531825\\
40	0.999885912531825\\
41	0.999885912531825\\
42	0.999885912531825\\
43	0.999885912531825\\
44	0.999885912531825\\
45	0.999885912531825\\
46	0.999885912531825\\
47	0.999885912531825\\
48	0.999885912531825\\
};
\addlegendentry{D-GW D-N ($m^*=3$)}

\addplot[const plot,line width=0.6mm, color=mycolor2] table[row sep=crcr] {%
1	0\\
2	0\\
3	0\\
4	0\\
5	0\\
6	0.0398898465217357\\
7	0.13458248057635\\
8	0.26947310147848\\
9	0.418947970336823\\
10	0.560944688108362\\
11	0.682369337957442\\
12	0.778529082485273\\
13	0.850343773484776\\
14	0.901550129823759\\
15	0.936717881758602\\
16	0.960133196351518\\
17	0.975322594440704\\
18	0.984959628301287\\
19	0.990958130794151\\
20	0.994630268629245\\
21	0.996845703299728\\
22	0.998165177722288\\
23	0.998942079921273\\
24	0.999394855798471\\
25	0.999656315597746\\
26	0.999806050565057\\
27	0.999891160396308\\
28	0.999939208249886\\
29	0.999966165177895\\
30	0.999981203644726\\
31	0.999989549728108\\
32	0.999994159668673\\
33	0.999996694855765\\
34	0.999998083461773\\
35	0.999998841236858\\
36	0.999999253352551\\
37	0.999999476776459\\
38	0.999999597551151\\
39	0.999999662661986\\
40	0.999999697676557\\
41	0.99999971646291\\
42	0.999999726520817\\
43	0.999999731894987\\
44	0.999999734761248\\
45	0.999999736287328\\
46	0.999999737098565\\
47	0.999999737529166\\
48	0.999999737757412\\
};
\addlegendentry{D-GW O-N ($m^*=5$)}

\addplot[const plot,line width=0.6mm, color=mycolor4] table[row sep=crcr] {%
1	0\\
2	0\\
3	0\\
4	0\\
5	0\\
6	0\\
7	0\\
8	0.0036738552984367\\
9	0.0175805292050303\\
10	0.0477312668756942\\
11	0.0968862513322731\\
12	0.163862853255988\\
13	0.244432817836268\\
14	0.332862439519714\\
15	0.423346068241006\\
16	0.510973617937244\\
17	0.592183180088309\\
18	0.664812645113835\\
19	0.727909170889921\\
20	0.781434879092049\\
21	0.825963368125516\\
22	0.862418771621332\\
23	0.891877100140488\\
24	0.915429908125938\\
25	0.934100555143585\\
26	0.948800166278523\\
27	0.960310943249216\\
28	0.969286679891228\\
29	0.976262945690994\\
30	0.981671778482958\\
31	0.985857621962079\\
32	0.989092622876451\\
33	0.991590335775262\\
34	0.993517471820603\\
35	0.995003671559055\\
36	0.996149461471424\\
37	0.997032631723067\\
38	0.997713290390038\\
39	0.998237835447929\\
40	0.998642057262335\\
41	0.998953551260599\\
42	0.999193588280797\\
43	0.99937856132273\\
44	0.999521102959947\\
45	0.999630947523453\\
46	0.999715595928479\\
47	0.999780828119473\\
48	0.999831097971742\\
};
\addlegendentry{O-GW O-N ($m^*=7$)}

\end{axis}
\end{tikzpicture}%
    \caption{{The CDF of the packet delay for the power control scheme at the optimal segmentation $m^*$.} }
     \label{cdf}
\end{figure}

  The detailed cumulative distribution functions (CDFs) of the delay when operating at the optimal segmentation for the three antenna schemes are shown in Fig.~\ref{cdf}. Due to packet queueing and transmission failures, the delay from packet generation to departure takes more than $m^*$ transmission cycles. The figure also highlights the significant delay reduction offered by directive antennas, which can be attributed to the higher throughput and less required segmentation. For instance, {$95\%$} of the packets are delivered within 10 Tx cycles in the D-GW D-N scheme. Such percentage reduces to around {$55\%$} and {$4\%$} in the D-GW O-N and O-GW O-N schemes, respectively.
 
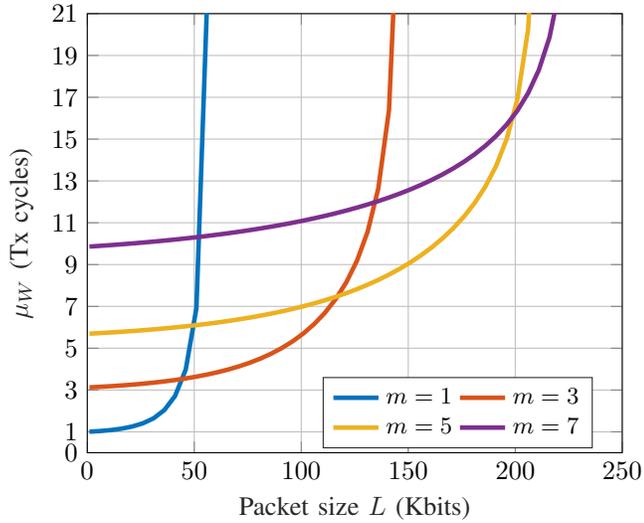
\begin{figure}[!h]
\centering
%
%
\definecolor{mycolor1}{rgb}{0.00000,0.44700,0.74100}%
\definecolor{mycolor2}{rgb}{0.85000,0.32500,0.09800}%
\definecolor{mycolor3}{rgb}{0.92900,0.69400,0.12500}%
\definecolor{mycolor4}{rgb}{0.49400,0.18400,0.55600}%
\definecolor{mycolor5}{rgb}{0.46600,0.67400,0.18800}%
\definecolor{mycolor6}{rgb}{0.30100,0.74500,0.93300}%
\definecolor{mycolor7}{rgb}{0.63500,0.07800,0.18400}%
\begin{tikzpicture}

\begin{axis}[%
width=2.8in,
height=2.3in,
at={(0.758in,0.481in)},
scale only axis,
xmin=0,
xmax=250,
xlabel style={font=\color{white!15!black}},
xlabel={Packet size $L$ (Kbits)},
ymin=0,
ymax=21,
ytick={0,1,3,...,21},
xmajorgrids,
ymajorgrids,
ylabel style={font=\color{white!15!black}},
ylabel={$\mu_W$ (Tx cycles)},
axis background/.style={fill=white},
legend style={font=\small,at={(0.44,0.01)}, anchor=south west, legend columns=2, legend cell align=left, align=left, draw=white!15!black}
]
\addplot [line width=0.6mm,  color=mycolor1]
  table[row sep=crcr]{%
1	1.00709719606135\\
6	1.03794672249769\\
11	1.08371448250435\\
16	1.15164896956446\\
21	1.25320103594364\\
26	1.40743691114682\\
31	1.64808152936031\\
36	2.03898590115088\\
41	2.71133633562977\\
46	3.97671324306039\\
51	6.91408861315647\\
56	21.5923933059025\\
};
\addlegendentry{$m=1$}

\addplot [line width=0.6mm,  color=mycolor2]
  table[row sep=crcr]{%
1	3.12682312801793\\
6	3.15328861947471\\
11	3.18350072754679\\
16	3.21797259850536\\
21	3.25728977479971\\
26	3.3021225997726\\
31	3.35324144613366\\
36	3.41153550546401\\
41	3.4780360792811\\
46	3.55394558026815\\
51	3.64067380909861\\
56	3.73988355116853\\
61	3.85354819207463\\
66	3.98402490864949\\
71	4.13414828680966\\
76	4.30735088877773\\
81	4.50782003174515\\
86	4.74070447805757\\
91	5.01239331572621\\
96	5.33090759936055\\
101	5.70648623664731\\
106	6.15253812075791\\
111	6.68732746295269\\
116	7.33718326133356\\
121	8.14303600787057\\
126	9.17492773842302\\
131	10.5687897804782\\
136	12.6410058606889\\
141	16.3883546167732\\
146	27.9184086062676\\
};
\addlegendentry{$m=3$}

\addplot [line width=0.6mm,  color=mycolor3]
  table[row sep=crcr]{%
1	5.69502075459188\\
6	5.72289189103557\\
11	5.75307241100231\\
16	5.78574699127959\\
21	5.821115231214\\
26	5.85939304471363\\
31	5.9008142698467\\
36	5.94563244541252\\
41	5.99412283818979\\
46	6.04658473585629\\
51	6.10334404915595\\
56	6.16475627035917\\
61	6.23120984302026\\
66	6.30313000780025\\
71	6.38098320133621\\
76	6.4652821007947\\
81	6.55659142740848\\
86	6.65553465045974\\
91	6.76280177285718\\
96	6.87915843710265\\
101	7.00545667630413\\
106	7.14264776503359\\
111	7.2917978240208\\
116	7.45410713844929\\
121	7.63093461809769\\
126	7.82382954240074\\
131	8.03457382116007\\
136	8.26523966629992\\
141	8.51827010501598\\
146	8.79659387319451\\
151	9.10379277067794\\
156	9.44435098331847\\
161	9.82403625509466\\
166	10.2505014561582\\
171	10.7342723155858\\
176	11.2904509045006\\
181	11.9418378937\\
186	12.7251057596833\\
191	13.7042411807923\\
196	15.0038118105327\\
201	16.9075772840623\\
206	20.2459145358079\\
211	29.0042255462849\\
216	391.364795355184\\
};
\addlegendentry{$m=5$}

\addplot [line width=0.6mm, color=mycolor4]
  table[row sep=crcr]{%
1	9.86068054808515\\
6	9.89476621390382\\
11	9.93086122101671\\
16	9.96908152706109\\
21	10.0095501473348\\
26	10.0523977134382\\
31	10.097763099764\\
36	10.1457941290648\\
41	10.1966483705827\\
46	10.2504940470031\\
51	10.3075110699523\\
56	10.367892228071\\
61	10.4318445571104\\
66	10.4995909283307\\
71	10.5713719079792\\
76	10.6474479008503\\
81	10.7281017482571\\
86	10.8136417443048\\
91	10.9044052716547\\
96	11.0007631680749\\
101	11.10312501603\\
106	11.2119455989728\\
111	11.3277328449331\\
116	11.4510576835608\\
121	11.5825663894344\\
126	11.722996190879\\
131	11.8731952182921\\
136	12.0341482933418\\
141	12.2070106904219\\
146	12.3931529474523\\
151	12.5942212516223\\
156	12.8122201876981\\
161	13.04962829092\\
166	13.3095627336835\\
171	13.5960198164702\\
176	13.9142354482371\\
181	14.2712424913179\\
186	14.6767636213822\\
191	15.1447026426908\\
196	15.6957628958142\\
201	16.3623321301735\\
206	17.1983125996297\\
211	18.3009301907572\\
216	19.8659061853782\\
221	22.3559398696465\\
226	27.1953836137941\\
231	41.9901462339201\\
};
\addlegendentry{$m=7$}


\end{axis}
\end{tikzpicture}%
    \caption{ { Average delay versus packet size for the D-GW D-N scheme. }}
     \label{data}
\end{figure}

The effects of packet size $L$ and the subsequent packet segmentation for the D-GW D-N scheme are shown in Fig. \ref{data}. The figure shows that larger packets have to be divided into more segments in order to be delivered with finite delay. However, over-segmenting packets leads to unnecessary delays. For instance, a packet of size 100 kbits can be divided into  {$m\in \{3,\;4,\cdots, 7\}$}, yet $m^*=3$ is the optimal value that minimizes delay. The figure also shows that there is a maximum packet size (e.g.,  {$L=220$} kbits in the figure) that cannot be delivered with finite delay. That is, going beyond  {7} segments will not enable the gateways to support packets larger than  {$K\!=\!220$} kbits with the temporal resolution of $T_a\!=\!18$~Tx~cycles. 
 
 Fig.~\ref{sta} illustrates the data granularity feasibility region for the different antenna schemes. The feasibility region is depicted by the shaded region in Fig.~\ref{sta}, which shows the data granularity parameters $(L,T_r)$ that can be supported with finite delay. The figure clearly demonstrates the contradicting tradeoff between the information content and temporal resolution. That is, more informative packets (i.e., with increased $L$) come at the expense of degraded temporal resolution (i.e., larger duration between packets) to maintain transmission feasibility. The figure also highlights the significant improvement in terms of data granularity that can be offered by directive antennas when compared to omni-antenna. For instance, for the temporal resolution of  {$T_r=12$ s}, the omni gateways and devices can support at most $L=50$ kbits packets. Equipping the gateways and devices with directive antennas enables handling 3 times larger packets. It is worth noting that the staircase-shaped boundary of the feasibility regions indicates the range of packet sizes $L$ that can be supported for each value of $T_r$. Furthermore, $T_r$ has to be increased by multiples of the Tx cycles $\mathcal{N}_G T_s$ in order to extend the feasibility region when increasing $L$.

\begin{figure}[!h]
\centering
  \input{stabilty}
    \caption{{Feasibility regions where finite delay is guaranteed.}}
     \label{sta}
\end{figure}

\section{Summary \& Conclusion}\label{conc}

This paper characterizes terrestrial data aggregation in large-scale regularly deployed IoT networks with synchronous time-triggered traffic. The periodic network traffic is parameterized by the sizes and inter-arrival times of the generated packets in order to account for the information content and temporal resolution of the aggregated data, respectively. The synchronously generated data in the IoT devices is aggregated via terrestrial gateways that adopt random scheduling and utilize universal frequency reuse. The IoT devices may partition large packets into smaller segments to maintain reliable transmission rate during the scheduled uplink time slots, which creates an intricate tradeoff between data granularity, transmission reliability, and data aggregation delay. Using stochastic geometry and queueing theory, we develop a novel spatiotemporal model to characterize such tradeoff while accounting for the potential utilization of directive antenna and channel inversion power control to improve and expedite the data aggregation process. 

To this end, we obtain accurate approximate expressions for the rate-aware and signal-to-interference-plus-noise-ratio (SINR) dependent segment transmission success probability for the different antenna and power control schemes. The periodic packet arrivals and the rate-aware and SINR-dependent packet departures are modeled via two different phases (PH) type distributions, which are utilized to construct and solve a novel PH/PH/1 queueing model to characterize the data aggregation delay. The numerical results reveal the necessity of packet segmentation and show the existence of an optimal number of segments that minimizes delay, which varies across the different antenna schemes. The results also quantify the significant performance improvement, in terms of reliability and delay, that can be achieved by implementing directive antennas. In contrast, path-loss power control is shown to degrade the network performance but provides a unified (i.e., location-independent) transmission success probability for all IoT devices. Interestingly, the performance degradation imposed by the power control is marginal when utilizing directive antennas. Finally, we characterize the joint feasible range of sizes and inter-arrival times of the data packets that can be aggregated within a finite time.


%

\appendices

\section{Proof of Lemma 1 }



\begin{figure}[!t]
\centering
  \includegraphics[width=0.4\textwidth]{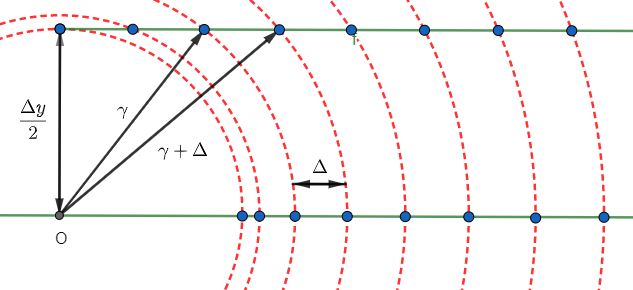}
    \caption{Compression due to projection for the first line (i = 0).}
    \label{compression}
\end{figure}

The homogeneous 1D-PPP of each line is projected into the central line  passing through the origin, that does not have any nodes. This forms a non-homogeneous 1D-PPP on a single line through a transformation of intensity (see \cite{farooq2015stochastic}) as depicted in Fig. \ref{compression}.
It is observed that the points experience a compression which is  function of the distance $\gamma$  between the point and the origin, where the compression factor is given by \eqref{comp}.
The intensity of the 1D-PPP of active nodes on the $i^{th}$ line, mapped to the line passing through origin is  $ C(\gamma,i)\frac{1}{\Delta x \mathcal{N}_G}$ $\mathrm{nodes/m}$. At the same time, this intensity of interferer nodes locations is marked by a mark of its antenna orientation and another mark of the antenna orientation of the test gateway at the origin w.r.t. it, to form a 3D-PPP with intensity  $\lambda_{T_1}$.

Laplace transform of the aggregate interference is
\small
\begin{equation}
\begin{aligned}[b]
\!\!&\mathcal{L}_{I_{agg}} (s)\! = \mathbb{E}_\Psi \left[\exp \left\{-s I_{agg}\right\}\right]  \\
&=\mathbb{E}_\Psi \left[\exp \left\{-s \sum_{i=1}^{\infty}\sum_{j\in\Psi_i}   \mathbbm{1}_{\{v_{ij} \text{ is active}\}}P h_{ij} v_{ij}^{-\eta}G(\theta_{T_{ij}})G(\theta_{R_{ij}})\right\}\right] \\
&= \displaystyle\prod_{i=1}^{\infty} \mathbb{E}_\Psi \left[   \displaystyle\prod_{j\in\Psi_i}    \exp \left\{-  \mathbbm{1}_{\{v_{ij} \text{ is active}\}}s   P h_{ij} v_{ij}^{-\eta} G(\theta_{T_{ij}})G(\theta_{R_{ij}})\right\}\right]
\end{aligned}
\end{equation}
\normalsize
 since $h_{ij}$ is unit mean exponential random variable, $h_{ij}\sim\text{Exp}(1)$, and from the definition of moment generating function \\*
\begin{equation}
\!\!\mathcal{L}_{I_{agg}} (s)\!= \displaystyle\prod_{i=1}^{\infty} E_\Psi \left[   \displaystyle\prod_{j\in\Psi_i} \frac{1}{1+ \mathbbm{1}_{\{v_{ij} \text{ is active}\}}sP v_{ij}^{-\eta}G(\theta_{T_{ij}})G(\theta_{R_{ij}})} \right]
\end{equation}
Using the new intensity of PPP over 3-D $\lambda_{T1}$ and from the definition of Probability Generating Functional (PGFL) of PPP
\begin{equation}
\begin{aligned}[b]
\!\!\mathcal{L}_{I_{agg}} (s)\!&= \exp \left\{- \sum\limits_{i=0}^{\infty} \int_{\theta_2}\int_{0}^{2\pi}\int_{\psi}\right.\\ &\left. \frac{4 \lambda_{T_1}}{1+\gamma^{\eta} (s P G(\theta_1)G(\theta_2))^{-1}} d\gamma d\theta_1 d\theta_2 \right\}
\end{aligned}
\end{equation}

 Due to the symmetry of interferer devices on both right and left halves  and the symmetry between the upper and lower halves of the network, the integration is performed over one quarter and  the intensity is multiplied by 4.\\
 By dividing the integrations over the two regions described in Approximation 1.
Where in first region, $\theta_2$ is from $\frac{\pi}{2}$ to $\frac{3 \pi}{2}$ and $\lambda_{T_1}=\slfrac{C(\gamma,i)\lambda_{1D}}{2\pi^2 }$, while the second region $\theta_2$ from $0$ to $2 \pi$ and $\lambda_{T_1}=\slfrac{C(\gamma,i)\lambda_{1D}}{(2\pi)^2}$.
Furthermore, substituting with interference region $\psi$ on each line as follow: for the lines that pass through the test cell, the argument $i$ takes the values from 0 to $Y$, and the interference exists from $\frac{\sqrt{3}R}{2}$ to $\infty$. While, the lines outside the test cell, $i$ takes the values from $Y+1$ to $\infty$ , and the interference  from $\frac{\Delta y}{2}(2i+1)$ to $\infty$. Hence, the lemma is proved.


\section{Proof of Lemma 2 }
Let The set $\Psi_j$ encompasses the locations  of all potential interfering devices of the grid network

\small
\begin{equation}
\begin{aligned}[b]
\!\!\mathcal{L}_{I_{agg}} (s)\!&=\mathbb{E}_\Psi \left[\exp\left\{-s I_{agg}\right\}\right]\\
&=E_\Psi \left[\exp\left\{-s \sum_{\Psi_j}  \mathbbm{1}_{\{v_{j} \text{ is active}\}}P h_{j} v_{j}^{-\eta}G(\theta_{T_{j}})G(\theta_{R_{j}})\right\}\right]\\
&=\mathbb{E}_\Psi \left[  \displaystyle\prod_{\Psi_j}   \exp\left\{- \mathbbm{1}_{\{v_{j} \text{ is active}\}}s   P h_{j} v_{j}^{-\eta}G(\theta_{T_{j}})G(\theta_{R_{j}})\right\}\right]
\end{aligned}
\end{equation}
\normalsize
since $h_{j}\sim\text{Exp}(1)$ and from the definition of moment generating function
\begin{equation}
\!\!\mathcal{L}_{I_{agg}} (s)\!= \mathbb{E}_\Psi \left[   \displaystyle\prod_{\Psi_j} \frac{1}{1+ \mathbbm{1}_{\{v_{j} \text{ is active}\}}sP v_{j}^{-\eta}G(\theta_{T_{j}})G(\theta_{R_{j}})} \right]
\end{equation}
By using polar coordinates to describe the 2-D interferers locations, and adding their antenna orientations, a new 3-D PPP of intensity  $\lambda_{T_2}$ is obtained. Utilizing the definition of PGFL of PPP, the Laplace transform of the aggregate interference can be expressed as
\begin{equation}
\begin{aligned}[b]
\!\!\mathcal{L}_{I_{agg}} (s)\!&
=\exp\left\{-\int_{\theta2}\int_{0}^{2\pi}\int_{\frac{\sqrt{3}R}{2}}^{\infty}\right.\\
&\left.\frac{v\lambda_{T_2}}{1+v^{\eta} (s P G(\theta_1)G(\theta_2))^{-1}} dvd\theta_1 d\theta_2\right\}
\end{aligned}
\end{equation}
By dividing the integrations over the two regions described in Approximation 1. Where in first region, $\theta_2$ is from $\frac{\pi}{2}$ to $\frac{3 \pi}{2}$ and $\lambda_{T_2}=\slfrac{\lambda_{2D}}{\pi}$, while the second region  $\theta_2$ from $0$ to $2 \pi$ and $\lambda_{T_2}=\slfrac{\lambda_{2D}}{2\pi}$, it can be shown that

 \begin{equation}\label{lap_area}
  \begin{aligned}[b]
\!\!\mathcal{L}_{I_{agg}} (s)\!&=\exp\left\{-\left[\int_{\pi/2}^{\frac{3 \pi}{2}} \int_{0}^{2\pi} \int_{\frac{\sqrt{3}R}{2}}^{\sqrt{3}R} H dvd\theta_1 d\theta_2 \quad + \right.\right.\\
&\left.\left.\int_{0}^{2\pi} \int_{0}^{2\pi} \int_{\sqrt{3}R }^{\infty} \frac{H}{2}  dvd\theta_1 d\theta_2\right]\right\}
  \end{aligned}
\end{equation}
\begin{equation}
H=\frac{\slfrac{\lambda_{2D} v}{\pi}}{1+v^{\eta} (s P G(\theta_1)G(\theta_2))^{-1}}
\end{equation}

Finally, by changing of variables the lemma is proved in terms of Gauss hypergeometric function $_2F_1$.

\section{proof of Lemma 3}
Let the set $\Psi_j$ encompasses the locations  of all potentially interfering devices of the grid network
\footnotesize
\begin{equation}
\begin{aligned}[b]
\!\!\mathcal{L}_{I_{agg}} (s)\!&=\mathbb{E}_\Psi \left[\exp\left\{-s I_{agg}\right\}\right]\\
&=\mathbb{E}_\Psi \left[\exp\left\{-\mathbbm{1}_{\{v_{j} \text{ is active}\}}s \sum_{\Psi_j}  P_j h_{j} v_{j}^{-\eta}G(\theta_{T_{j}})G(\theta_{R_{j}})\right\}\right]\\
&=\mathbb{E}_\Psi \left[  \displaystyle\prod_{\Psi_j}  E_{P_j,h_j} \left[\exp\left\{-\mathbbm{1}_{\{v_{j} \text{ is active}\}}s   P_j h_{j} v_{j}^{-\eta}G(\theta_{T_{j}})G(\theta_{R_{j}})\right\}\right]\right]
\end{aligned}
\end{equation}
\normalsize
since $h_{j}\sim\text{Exp}(1)$ and from the definition of moment generating function \\*
\begin{equation}
\!\!\mathcal{L}_{I_{agg}} (s)\!= \mathbb{E}_\Psi \left[   \displaystyle\prod_{\Psi_j} \mathbb{E}_{P_j} \left[\frac{1}{1+\mathbbm{1}_{\{v_{j} \text{ is active}\}}sP_j v_{j}^{-\eta}G(\theta_{T_{j}})G(\theta_{R_{j}})} \right]\right]
\end{equation}
Similar to the proof of Lemma 2, by using polar coordinates  and the  intensity of the PPP over 3-D $\lambda_{T_2}$ and from the definition of PGFL of PPP 
\begin{equation}
  \begin{split}
\!\!\mathcal{L}_{I_{agg}} (s)\!&= \exp\left\{-\int_{\theta_2}\int_{0}^{2\pi}\int_{\frac{\sqrt{3}R}{2}}^{\infty} \lambda_{T_2} \right.\\
&\left. \mathbb{E}_{p_j} \left(1- \frac{1}{1+sP_j v^{-\eta}G(\theta_2)G(\theta_1)} \right)vdv d\theta_1 d\theta_2\right\}
\end{split}
\end{equation}
By dividing the integrations over the two regions described in Approximation 1. Where in first region, $\theta_2$ is from $\frac{\pi}{2}$ to $\frac{3 \pi}{2}$ and $\lambda_{T_2}=\slfrac{\lambda_{2D}}{\pi}$, while the second region  $\theta_2$ from $0$ to $2 \pi$ and $\lambda_{T_2}=\slfrac{\lambda_{2D}}{2\pi}$, then making change of variables of $y=\frac{v}{\left(sP_j G(\theta_2)G(\theta_1)\right)^{1/\eta}}$.
Noting that an interferer device at $\frac{\sqrt{3}R}{2}$ from test gateway is transmitting power of $\rho (\frac{\sqrt{3}R}{2})^{\eta}$ to its gateway. Also, an interferer device at $\sqrt{3}R$ from test gateway is transmitting power of$\rho (\frac{\sqrt{3}R}{3})^{\eta}$ to its gateway, and $\mathbb{E}_{P_j}\left[P_j^{2/\eta}\right]=\rho^{2/\eta}\mathbb{E}\left[r^2\right]$ where  $r$ is the distance between the interfering device and its gateway.
It can be shown that
\small
\begin{equation}\label{lap_area_pc}
  \begin{aligned}[b]
&\mathcal{L}_{I_{agg}} (s)\!=\exp\left\{- s^{2/\eta} \rho^{2/\eta}\mathbb{E}\left[r^2\right]\right.\\
&\left.\left[\int\limits_{\pi/2}^{\frac{3 \pi}{2}}\int\limits_{0}^{2\pi}\int\limits_{B}^{3 B} \frac{\lambda_{2D} F}{\pi}  dy d\theta_1 d\theta_2
+ \int\limits_{0}^{2\pi}\int\limits_{0}^{2\pi}\int\limits_{3 B}^{\infty} \frac{\lambda_{2D} F}{2\pi}  dy d\theta_1 d\theta_2\right]\right\}
  \end{aligned}
\end{equation}
\normalsize
where
\begin{equation}
\begin{aligned}[b]
B&=\left(s \rho G(\theta_2)G(\theta_1)\right)^{-1/\eta}\\
F&=\frac{y}{y^\eta +1} \left( G(\theta_2)G(\theta_1)\right)^{2/\eta} \nonumber
\end{aligned}
\end{equation}
which can be further simplified using Gauss hypergeometric function $_2F_1(\cdot)$, and hence the lemma is proved.





\section*{Acknowledgment}
This work was supported by King Fahd University of Petroleum and Minerals (KFUPM), Dhahran, Saudi Arabia, under project No GTEC1802.

\ifCLASSOPTIONcaptionsoff
  \newpage
\fi



%
\bibliographystyle{ieeetr}
\bibliography{IoT_paper.bib}

%








\end{document}